\documentclass{emulateapj}
\usepackage{amsmath,natbib}

\def\lt{{\lambda}'}

\def \beq {\begin{equation}}
\def \eeq {\end{equation}}
\def \bea {\begin{eqnarray}}
\def \eea {\end{eqnarray}}

\shorttitle{The cosmic history of the spin of dark matter haloes}
\shortauthors{Trowland, Lewis, Bland-Hawthorn}

\begin{document}

\renewcommand{\thefootnote}{\fnsymbol{footnote}}
\title{The cosmic history of the spin of dark matter haloes within the large scale structure \footnotemark[1]}

\author{Holly E. Trowland, Geraint F. Lewis and Joss Bland-Hawthorn}
\affil{Sydney Institute for Astronomy, School of Physics A28, The University of Sydney, NSW 2006, Australia}
\email{h.trowland@physics.usyd.edu.au}

\begin{abstract}

We use N-body simulations to investigate the evolution of the orientation and magnitude of dark matter halo angular momentum within the large scale structure since z=3. We look at the evolution of the alignment of halo spins with filaments and with each other, as well as the spin parameter, which is a measure of the magnitude of angular momentum. It was found that the angular momentum vectors of dark matter haloes at high redshift have a weak tendency to be orthogonal to filaments and high mass haloes have a stronger orthogonal alignment than low mass haloes. Since z=1, the spins of low mass haloes have become weakly aligned parallel to filaments, whereas high mass haloes kept their orthogonal alignment. This recent parallel alignment of low mass haloes casts doubt on tidal torque theory as the sole mechanism for the build up of angular momentum. We see evidence for bulk flows and the broadening of filaments over time in the alignments of halo spin and velocities. We find a significant alignment of the spin of neighboring dark matter haloes only at very small separations, $r<0.3$Mpc/h, which is driven by substructure. A correlation of the spin parameter with halo mass is confirmed at high redshift.

\end{abstract}


\footnotetext[1]{Research undertaken as part of the Commonwealth Cosmology Initiative (CCI: www.thecci.org), an international collaboration supported by the Australian Research Council}


\section{Introduction}

The large scale structure of the universe observed today has formed by a long history of gravitational collapse, gradual accretion and mergers. Through these processes a filamentary, sponge-like structure has emerged. The distribution of galaxies and their motions provides clues on how they formed, and together with galactic angular momentum data, the emergence of the intricate large scale structure can begin to be explained.

Before we can determine what spin tells us about the formation of large scale structure, the mechanisms of angular momentum build-up need to be well understood. The initial spin of early dark matter proto-haloes can be predicted analytically \citep{white84}, however these predictions are largely limited to the regime of linear structure formation. To track the angular momentum build up through more recent cosmic history, N-body simulations of cold dark matter must be used. These simulations give full information on the dark matter haloes which can be used to form hypothesis on the build up of galaxy angular momentum on cosmological scales. However, on cosmological scales it is not yet feasible to simulate the gas component to track the angular momentum build up of galaxies directly (although \citet{2010MNRAS.405..274H} simulated 100 disk galaxies in a filament, to find an alignment of galaxy spin with filaments).

Hydrodynamical simulations on individual galaxy scales \citep{2003MNRAS.346..177V, 2005ApJ...628...21S, 2010MNRAS.404.1137B} have shown that the specific angular momentum of baryons remains close to that of dark matter and that the galaxy angular momentum is generally about $20^\circ$ misaligned with the dark matter halo. This means that dark matter halo spin is a fairly good proxy for galaxy spin, so some understanding of the spins of galaxies may be gleamed from dark matter-only simulations. The spin of a dark matter halo depends mainly on two things; the initial torques driven by the surrounding landscape at early times, and the accretion and merger history of the halo. 

The initial spin of dark matter haloes is given through a mechanism known as ``tidal torque theory'', pioneered by \citet{hoyle}, \citet{peebles69} and \citet{Zel'Dovich}. This theory proposes that the initial spin of a proto-halo early in its formation in the linear regime of structure formation depends on its shape and the tidal forces exerted from the surrounding structure, so the spin is dependent on the local dark matter landscape. The greatest effects of tidal torquing happen at the time of turn-around, just before the proto-haloes have collapsed to virialized objects. A halo that was torqued in this manner should retain some memory of the tidal field where it formed, and this has been confirmed through N-body simulations and galaxy catalogues \citep[eg.][]{2001astro.ph.11186L, 2002MNRAS.332..325P, 2007ApJ...671.1248L}. The cosmic web is the manifestation of the tidal field, filaments in particular are regular, symmetric morphologies which on large scales exhibit a uniform tidal field. Thus it is expected that the orientation of halo spin today should retain some correlation with the direction of filaments and haloes should be aligned with each other over short distances.

Since the epoch of tidal torquing, halo spins have been substantially influenced by mergers and accretion. It was shown in \citet{2011arXiv1104.0935B} that it is not uncommon for the direction of the spin of a halo to completely flip over in its lifetime and this phenomenon is caused by minor and major mergers and even close halo flybys. Satellite accretion has been proposed to be the main contributor of angular momentum  and it has been shown that by neglecting tidal torques and considering mergers alone, the distribution of the magnitude of spin can be reproduced \citep[see][]{2001ApJ...557..616G, 2002ApJ...581..799V, 2002MNRAS.329..423M}.

To figure out how accretion has influenced dark matter halo spin and what spin can reveal about the formation of large scale structure, several authors have investigated an alignment of spin with the cosmic web using N-body simulations and galaxy catalogues. 
In simulations, it has been found that spins are aligned on shells around voids, lying preferentially on the void surface \citep{2007MNRAS.375..184B, 2008MNRAS.385..867C}. It has been shown that spins lie preferentially in the plane of sheets in simulations \citep{2004ApJ...613L..41N} and along the axis of filaments \citep{2002A&A...395....1F, 2007ApJ...655L...5A, 2007MNRAS.375..489H, 2009ApJ...706..747Z}. In observations there has been a tentative detection of some weak correlation with filaments \citep{2010MNRAS.408..897J} but no significant detection has been found to date.
The evolution of halo spin with respect to filaments and sheets was explored by \citet{2007MNRAS.381...41H} who found no change in the orientation of spin over cosmic time.

Since the spins of haloes are aligned with the large scale structure, there should be some degree of coherence between the direction of spin of two neighbouring haloes. It is not clear if this alignment is strong enough to be detected even in N-body simulations. \citet{2000MNRAS.319..649H}, \citet{2002MNRAS.332..325P}, \citet{2002A&A...395....1F}   and \citet{2005ApJ...627..647B} see no strong alignment whereas \citet{2001MNRAS.322..576H} do see a significant alignment. In contrast, several claims have been made of spiral galaxy spin alignments in observations \citep{2000ApJ...543L.107P, 2009MNRAS.392.1225S, 2011ApJ...732...99L}. If these alignments can be seen in observations but not in dark matter simulations then it is a possible indication that the spins of the luminous galaxies are not aligned with their dark matter haloes. 

As well as the orientation, the magnitude of the spin may reveal secrets of the large scale structure. The spin parameter is a dimensionless measure of the amount of rotation of a dark matter halo and it has been found \citep{lemson,2008MNRAS.388..863C} not to depend on cosmology or environment. Both \citet{2008ApJ...678..621K} and \citet{2011MNRAS.411..584M} find a mass dependence of the spin parameter at high redshift but not at low redshift.

Observations of galaxy spin alignments in the large scale structure to date have only been through inferred galaxy spin orientations from observed disk galaxy shape. For example, \citet{2007ApJ...671.1248L} used the Tully catalogue of nearby spirals \citep{1974UppOR...5....0N, 1982euse.book.....L} to infer spin from the axial ratio (to find an alignment with the tidal field) and \citet{2009MNRAS.392.1225S} used the apparent sense of spiral rotation in the Galaxy Zoo catalogue. Direct measurements of galaxy rotation have been done with integrated field units (IFU) although only one galaxy is targeted at a time and it is not feasible to conduct a survey of large scale structure with direct spin measurements. However, a new multi-object IFU instrument has been developed which will enable a survey of $10^{4-5}$ galaxies in a volume limited sample \citep{2011OExpr..19.2649B,Croom}. There will soon be a huge influx of galaxy spin data, which has never been sampled before in such high volumes. In order to get the most out of these data and to direct future surveys, the dark matter halo spin must be better understood.


Our paper is organized as follows. First, the method are described in Section \ref{method}. Here we describe the set of simulations used in Section \ref{nbody}, then we discuss the characteristic mass scale for halo collapse in Section \ref{Mchar} and the method used for finding features in the large scale structure are described in  Section \ref{quantifying}. Theoretical predictions from Tidal Torque theory are discussed in Section \ref{alignment} and the results of alignment of halo spin with filaments and the alignment of neighbouring haloes' spins are presented. Results of the evolution of the spin parameter in are presented in Section \ref{SPSection}. Lastly we summarize and discuss our results in Section \ref{summary}

\section{Method} \label{method}
\begin{figure*}
\begin{minipage}[l]{7in}
\plottwo{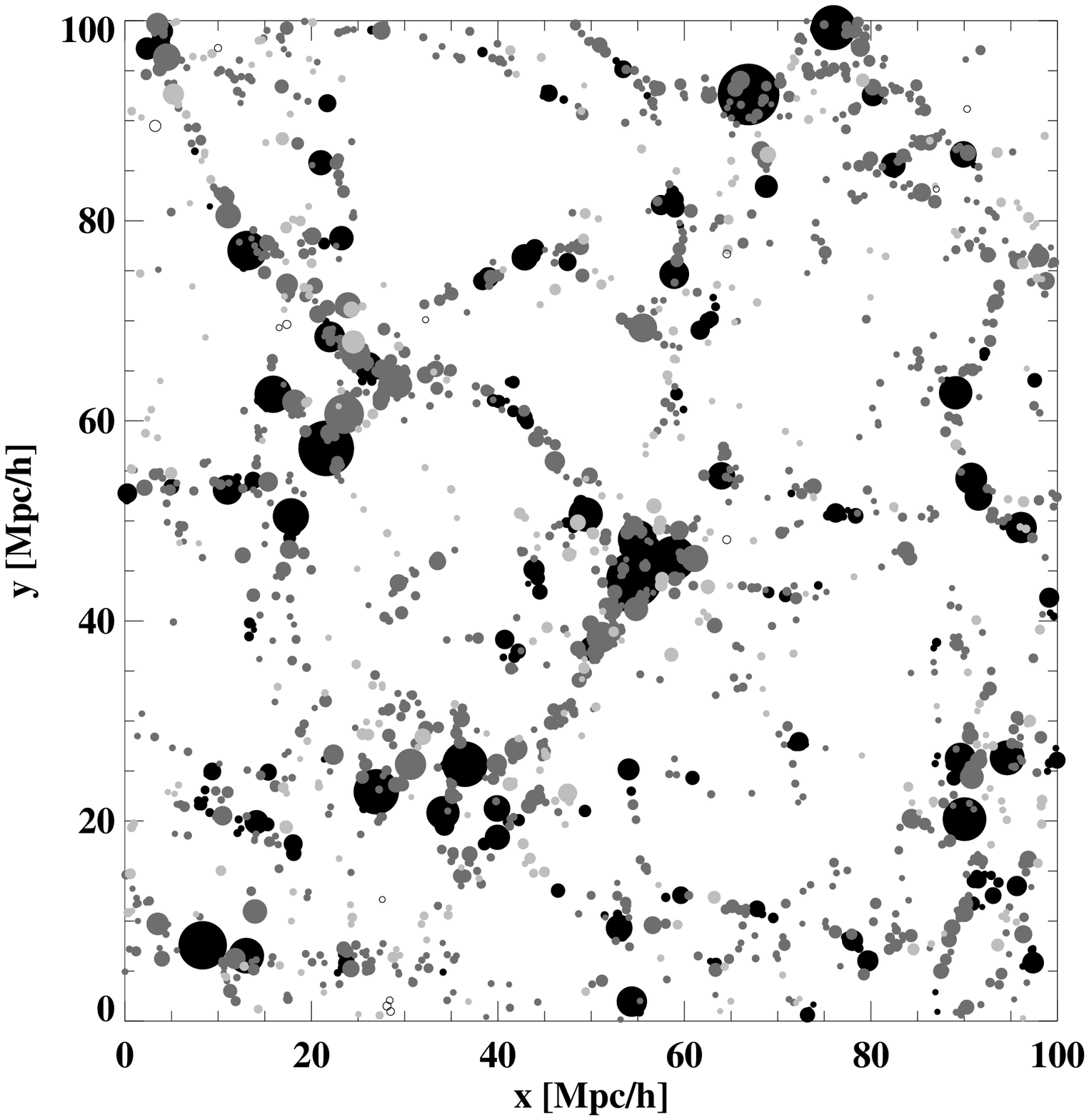}{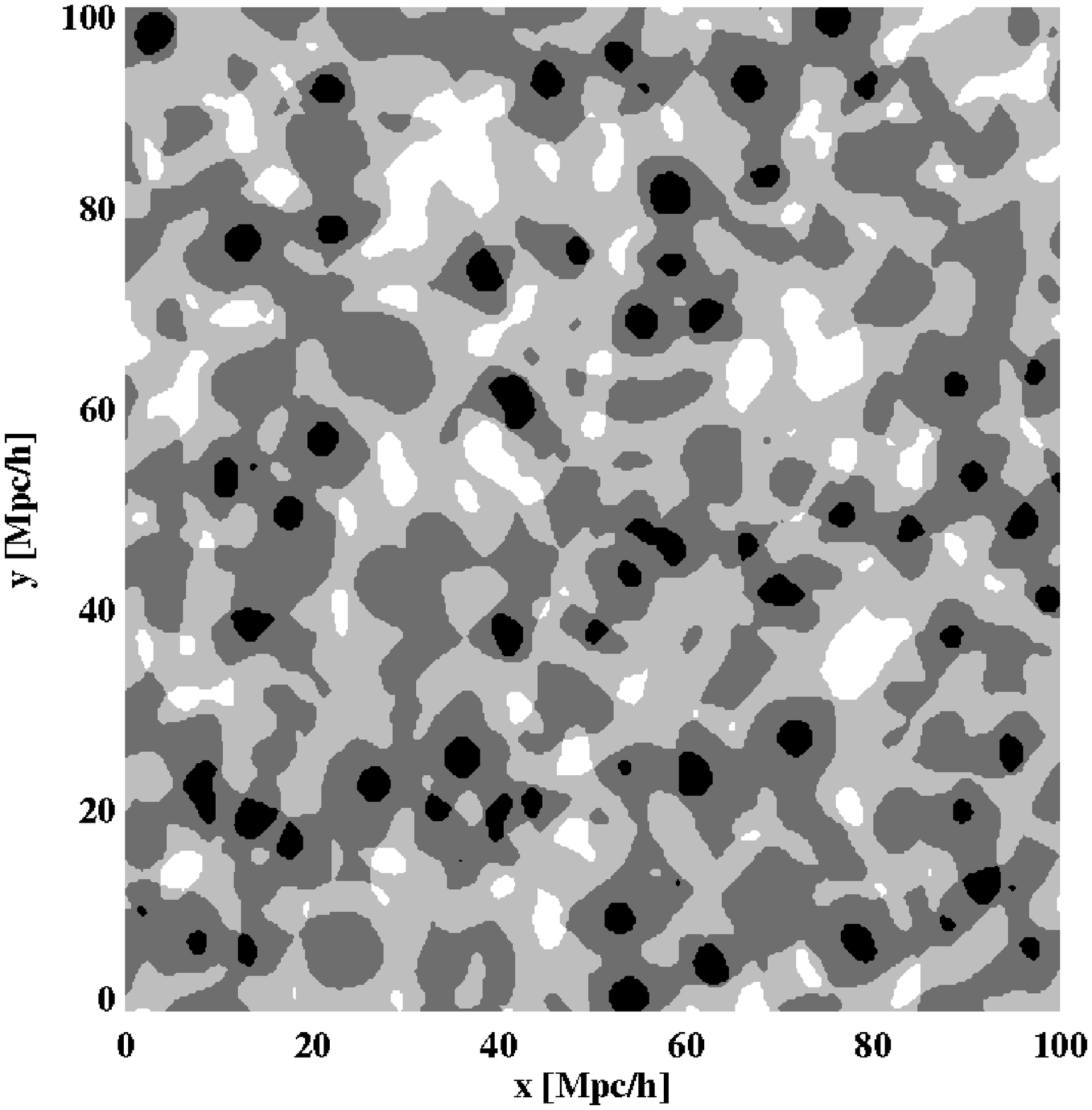}
\end{minipage}
\caption{\textbf{Left}: The distribution of dark matter haloes in a volume of the simulation where the large scale structure has been dissected into its component features. Haloes in blob regions are colored black, filament haloes are dark gray, sheet haloes are light gray and haloes in voids are outlined in black. The size of the dots are proportional to the virial radius of the halo and the volume shown is $100\times100\times5$ Mpc/h.
\textbf{Right}: The volume of the simulation is uniquely classified into features of the large scale structure using the dark matter density field. Here the classification of the volume is shown through the shading: blob regions are black, filaments are dark gray, sheets are light gray and voids are white. The features have been found on the scale of 2Mpc/h. \label{HaloSlice}}
\end{figure*}

\subsection{N-Body Simulation} \label{nbody}
Since any relic alignments of spin with the large scale structure are expected to be weak, a large simulation volume and high resolution are needed. To this end, the publically available Millennium simulation of \citet{2005Natur.435..629S} was used. This simulation is of a cubic volume $500\, $Mpc/h on a side containing $2160^3$ particles using the GADGET-2 code \citep{gadget2}. This gives a particle mass of $8.6\times 10^8 M_\odot$/h. A $\Lambda$CDM cosmology is chosen and the parameters are $\Omega_m=0.25$, $\Omega_b=0.045$, $\Omega_\Lambda=0.75$, h$=0.73$, $n=1$ and $\sigma_8=0.9$. 

The halo catalogue was built by \citet{2005Natur.435..629S} by first using the simple friends-of-friends group (FOF) finder \citep{fof} to attempt to select structure in the particle distribution and then finding the virialized subhaloes within the FOF groups using SUBFIND \citep{2001MNRAS.328..726S}. The SUBFIND algorithm first identifies subhalo candidates within each FOF halo using dark matter density and then removed particles that are not gravitationally bound to the subhalo candidate. The most massive subhalo typically contains most of the mass of the corresponding FOF object, and so can be regarded as the selfbound background halo itself, with the remaining subhaloes as its substructure. The halo catalogue used in this paper includes all virialized haloes, including subhaloes, although spin measurements are only made on haloes with more than 500 particles in order to minimize random effects from outer halo particles. There are 184,891 FOF haloes and 213,799 haloes in total.

For this analysis, a $300\, $Mpc/h section of the full Millennium simulation was used. This smaller section was chosen so that the resolution of the density field was high enough to be able to find features in the large scale structure. This was tested using several $100\,$Mpc/h sample cubes. As the resolution of the density field was raised from $128^3$ to $1024^3$ voxels, the alignment between halo spin and the resulting filaments became stable above a certain threshold. For smoothing lengths 2.0, 3.5 and 5.0 Mpc/h (Gaussian smoothing is used for finding filaments on different scales, see Section \ref{quantifying}), the minimum resolution for stable features is $0.4\,$Mpc/cell.  For a grid of $1024^3$ voxels, the maximum box size is $400\,$Mpc/h. To ensure the resolution was more than sufficient, a box of size $300\,$Mpc/h was chosen.  For smoothing on $1.0\,$Mpc/h scales, a finer grid must be used and the maximum cell size is $0.2\,$Mpc so a $200\,$Mpc/h box was used for this scale. At smaller scales than $1\,$Mpc/h the boxsize required is too small so there are not enough haloes for useful results. The following results display no cosmic variance when a different sample of the same size is chosen. There are 4,027,242 haloes in our $300\,$Mpc/h box and 932,961 haloes with more than 500 particles from which a reliable spin measurement could be made. The haloes in a 5Mpc slice through the simulation volume are shown in Figure \ref{HaloSlice}.


Snapshots are taken at several points throughout the simulation. Here we have used the snapshots at redshift 0, 0.99, 2.07 and 3.06 (rounded to 0, 1, 2, 3).

\subsection{Characteristic Mass}\label{Mchar}

In structure formation, there is a characteristic mass scale for collapse, $M_\ast(z)$. A spherical top-hat perturbation collapses when its linear overdensity exceeds a value of $\delta_c=1.686$. The variance of linear density fluctuations at a given mass scale $M$ is related to the linear power spectrum $P(k,z)$ at redshift $z$ by
\begin{equation}
\sigma^2(M,z) = \frac{1}{2\pi^2}\int_0^{\infty}dk\,k^2\,P(k,z)\,\widetilde{W}^{2}_{\rm{TH}}(k,M),
\end{equation}
where $\widetilde{W}_{\rm{TH}}(k,M)$ is the Fourier transform of a spherical top-hat window function of comoving size $R=(3M\,/\,4\pi\bar{\rho})^{1/3}$, and $\bar{\rho}$ is the comoving mean mass density of the universe.  At a given redshift, the typical mass scale $M_\ast(z)$ to collapse from a $1\sigma$ fluctuation is hence given by the implicit solution of 
\begin{equation}
\sigma(M_\ast,z)=\delta_c.
\end{equation}

The calculated values of characteristic mass at redshift 0, 1, 2 and 3 are 5.89, 0.273, 0.0132, $4\times10^{-5}$, respectively in units of $10^{12}M_\odot$.




\subsection{Quantifying the large scale structure}\label{quantifying}

Morphological features in large scale structure may be classified into four general categories: blobs, filaments, sheets and voids. This analysis uses the curvature of the density field to identify each of these features in N-body simulations.

Firstly, the density field is obtained using the Delaunay Tessellation Field Estimator (DTFE) method using the dark matter halo distribution (see \citet{CosmicWeb,DTFE,SchaapThesis}). The DTFE method can be summarized in three steps, i) from the distribution of points the Delaunay tessellation is constructed, which is a volume covering division of space into mutually distinct Delaunay tetrahedra. A Delaunay tetrahedron is defined by the set of four points whose circubscribing sphere does not contain any of the other points in the generating set. ii) The local density at each point is calculated from the volume of the Voronoi cells (the dual of the Delaunay tessellation) and the mass of the contained halo. iii) The density within each Voronoi cell is interpolated, assuming the density field varies linearly. The DTFE method is useful when looking for geometrical features in the density field because it automatically adapts to variations in density and geometry. 

The DTFE was carried out with vacuum boundary conditions and a buffer region around the box. This buffer region was made to be at least as big as the maximum distance between nearest neighbor haloes so that no voronoi cells constructed leaked outside the filled region. For larger smoothing scales, the buffer was at least as big as $2\sigma$. For the 2 and 3.5 Mpc scales the buffer was 7Mpc and for the 5Mpc scale the buffer was 10.5Mpc. The buffer region was also used in the smoothing of the density field then discarded.

Smoothing the density field to some scale $s$ is done by convolving with a spherically symmetric Gaussian filter,
\begin{equation}\label{convolution}
\rho_s(\textbf{x})=\int{d\textbf{y}\rho(\textbf{y})G_s(\textbf{x,y})}.
\end{equation}
Here $\rho(\textbf{y})$ is the Fourier transform of the DTFE density and the Gaussian filter at scale $s$ is defined by,
\begin{equation}\label{gaussian}
 G_s=\frac{1}{(2\pi \sigma_s^2)^{3/2}}\exp\left(-\frac{(y-x)^2}{2\sigma_s^2}\right)
\end{equation}
The curvature of the density field is given by the Hessian matrix of second derivatives at each point,
\begin{equation}\label{hessian}
 H_{\alpha\beta}=\frac{\partial^2\rho_s(\textbf{x})}{\partial x_\alpha\partial x_\beta}
\end{equation}
The second derivatives can be found while simultaneously smoothing the field by making use of an identity of the convolution; $\frac{d}{dx}(f*g)=\frac{df}{dx}*g=f*\frac{dg}{dx}$. Applying this to Equation \ref{convolution} gives
\begin{equation}
\frac{\partial^2\rho_s(\textbf{x})}{\partial x_\alpha\partial x_\beta}=\int{d\textbf{y}\rho(\textbf{y})  \frac{\partial^2}{\partial x_\alpha\partial x_\beta} G_s(\textbf{x,y})}.
\end{equation}
Thus, the Hessian of the smoothed density field is simply given by the convolution of the DTFE density and the second derivative of the Gaussian (the so-called `Mexican Hat wavelet'.)
\begin{equation}
 H_{\alpha\beta}=\frac{1}{\sigma_s^4}\int d\textbf{y} \rho(\textbf{y})[(x_\alpha-y_\alpha)(x_\beta-y_\beta)-\delta_{\alpha\beta}\sigma_s^2]G_s
\end{equation}
The eigenvalues of the Hessian quantify the curvature of density at a particular point, in the direction of the corresponding eigenvector. A positive eigenvalue indicates that the shape of the density field is concave up and a negative is concave down. The density field may now be classified uniquely into blob, filament, sheet or void regions according to the eigenvalues of this Hessian. The eigenvalue sign criteria for each region is as follows,
 
\begin{quote}
\begin{tabular}{l@{\quad}p{1.8in}}
\bf Blob & All negative\\
\bf Filament & Two negative, one positive\\
\bf Sheet & Two positive, one negative\\
\bf Void & All positive
\end{tabular}
\end{quote}

It can be useful to classify every point into one of these features as was done in \cite{2009ApJ...706..747Z}, and an alternative approach is to pick out only the best features like in \cite{MMF}. The decomposition of volume into features is shown in Figure \ref{HaloSlice} on the scale of 2Mpc/h. The filament and sheet morphologies dominate the volume, with blob regions taking up the least volume. The relative volume fractions do not change much over scale.

Morphological features are defined using only the eigenvalues of the Hessian. The direction of the eigenvectors are also used to assign a directionality to filaments and sheets. The direction of the axis of a filament is the direction of the positive eigenvalue, and the normal direction of a sheet is the direction of the negative eigenvalue. The features discussed in this paper have been found choosing the smoothing scales of 2.0, 3.5 and 5.0 Mpc/h. These scales have been chosen to match with the visual classification of structure at 2Mph/h \citep{2007MNRAS.375..489H} and to explore the scales above that. The comoving smoothing scales are kept constant for different redshifts in order not to bias the results with preconceived assumptions about filament formation.

This feature finding algorithm uniquely identifies regions into blob, filament, sheet or void depending only on the scale and quality of features required.

\section{Alignment of halo spin with the cosmic web} \label{alignment}

\begin{figure}  
\centering
\includegraphics[width=2.9in,height=2.9in]{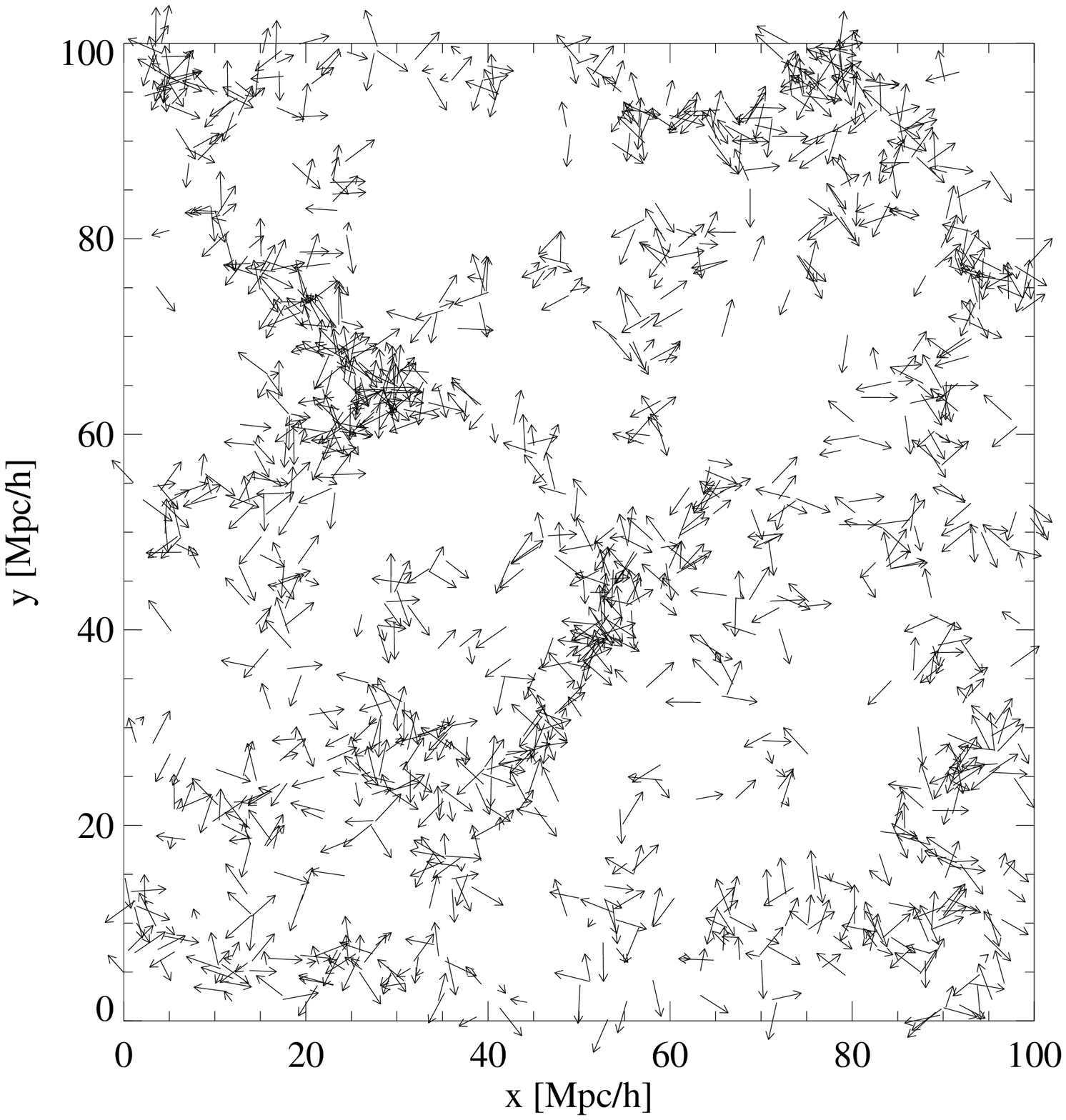}
\includegraphics[width=2.9in,height=2.9in]{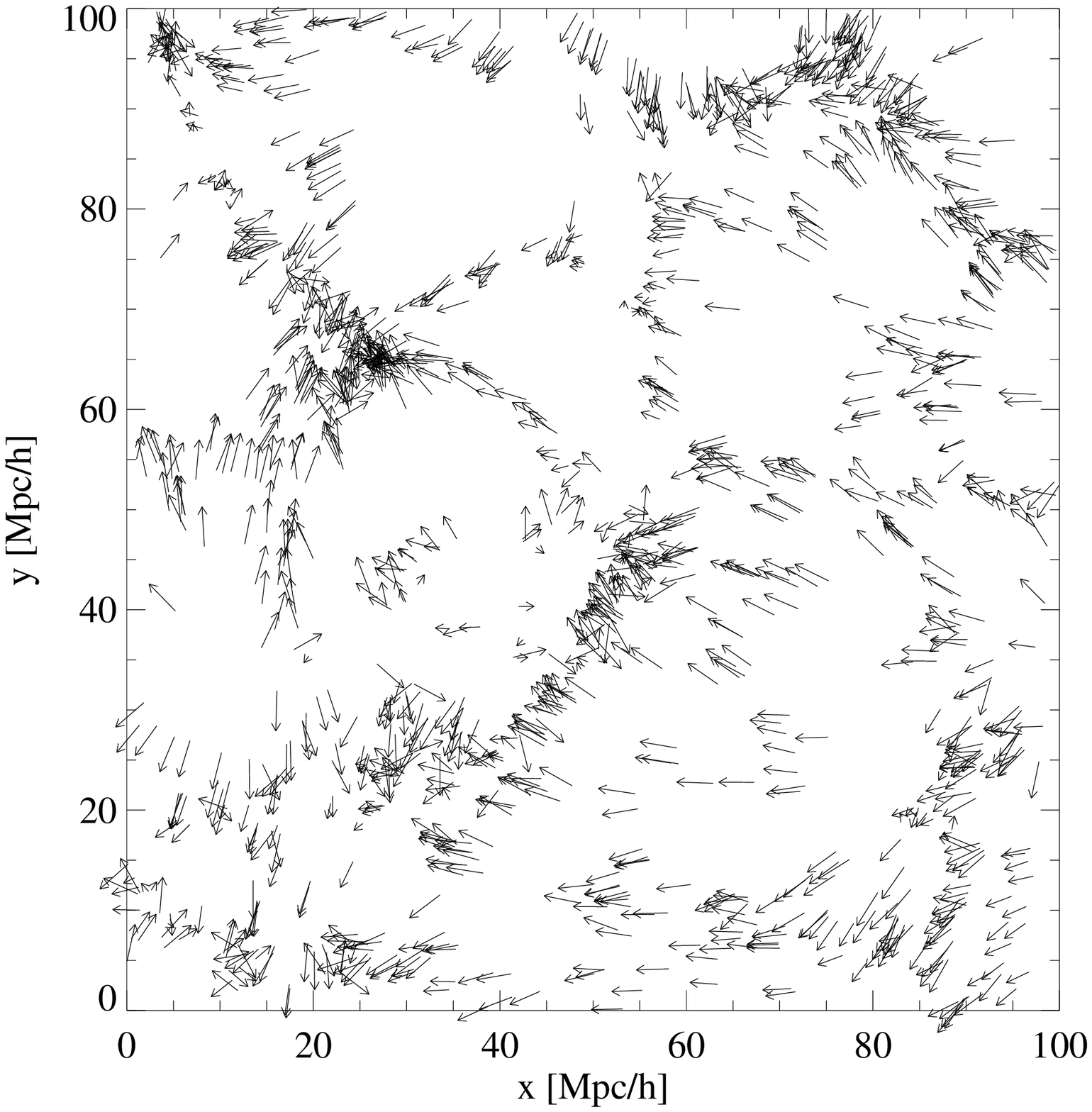}
\includegraphics[width=2.9in,height=2.9in]{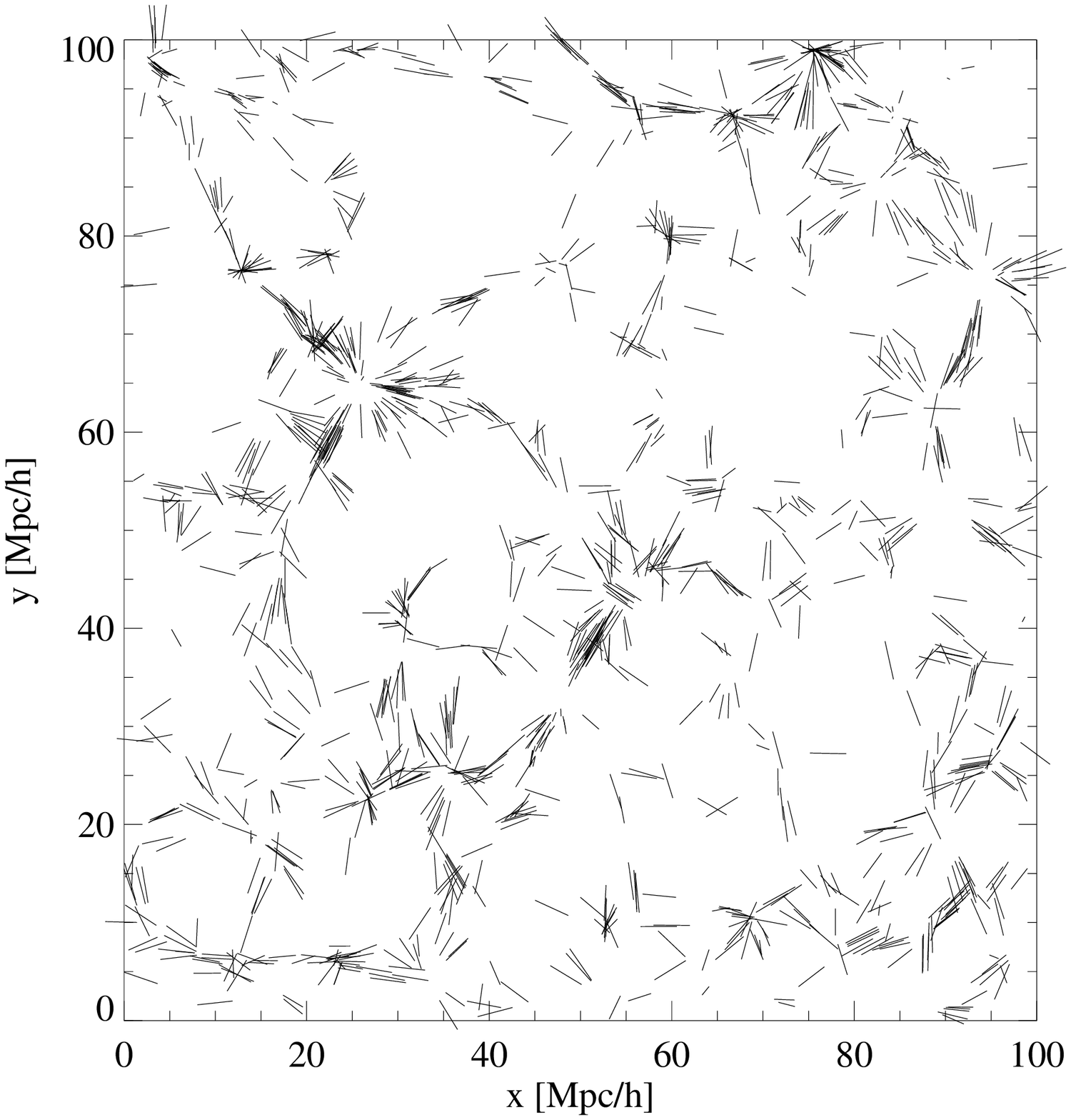}
\caption{ The direction of dark matter halo spin vectors (top), velocity vectors (middle) and filament axis (bottom). The velocities show a coherent flow along filament axis whereas spin vectors are much more random and not obviously aligned. Shown is a slice of the simulation $100\times100\times5$ Mpc/h and all vectors have been normalized to have the same length. \label{vectors}}
\end{figure}

Halo particles can be loosely bound, following stochastic paths, but adding up each particles angular momentum gives the nett effect of a halo spin. Spin is calculated by adding up the angular momentum of each particle ($i$) in the halo, simply defined as the cross product of the distance of the particle from the halo's centre of mass ($\bf r$) and the particles velocity ($\bf v$) with respect to the centre of mass;
\begin{equation}
{\bf J}=\displaystyle\sum\limits_{i=0}^N {\bf r}_i\times m_i {\bf v}_i
\end{equation}
In order to get a reliable measurement of halo spin, only the haloes with more than 500 particles have been included. The unit spin vectors are shown in the top panel of Figure \ref{vectors} but there is no obvious alignments with each other or with the large scale structure (as defined by the axis of filaments, shown on the bottom panel).

From tidal torque theory (TTT), the spin of dark matter haloes is expected to be correlated with the local tidal field (${\bf T}=T_{ij}\equiv \partial_i\partial_j\phi$) and the inertia tensor (${\bf I}=I_{ij}$). During the linear regime (assuming that $\bf{T}$ and $\bf{I}$ are uncorrelated), the first order result from TTT \citep{white84} is,
\begin{equation}
J_i\propto \epsilon_{ijk}T_{jl}I_{lk}, \label{TTTeqn}
\end{equation}
where $\epsilon_{ijk}$ is the Levi-Civita symbol. In the principle axis frame of the tidal tensor, where $\lambda_i$ are the eigenvalues of the tidal field, 
\begin{eqnarray*}
J_1\propto(\lambda_2-\lambda_3)I_{23}\\
J_2\propto(\lambda_3-\lambda_1)I_{31}\\
J_3\propto(\lambda_1-\lambda_2)I_{12} 
\end{eqnarray*}
$\lambda_3\leq\lambda_2\leq\lambda_1$ so $\lambda_3-\lambda_1$ is the largest coefficient, making $J_2$ the largest component of $\bf{J}$ so that spin is preferentially aligned with the second eigenvector of the tidal field. The cosmic web is a manifestation of the potential $\phi$, related by the Poisson equation, $\bigtriangledown^2\phi=4\pi G\rho\,(\textbf{x})$. Our definition of a filament (having two negative eigenvectors of the Hessian of density) translates into a region where there are two positive eigenvectors of the tidal tensor. The second eigenvector of the tidal field points in a direction orthogonal to the filament (the minor axis of the tidal field is the axis of the filament) and so we expect that halo spin should point in a direction orthogonal to the axis of the filament.

The result from TTT in Equation \ref{TTTeqn} assumes that $\bf{T}$ and $\bf{I}$ are completely uncorrelated, which has been shown to be not always true  \citep{ 2000ApJ...532L...5L,2002MNRAS.332..325P}. If there is some correlation, the preferred direction of halo spins discussed above may be a small effect. The alignment would also be greatly affected by merger and accretion events that have happened during nonlinear structure growth. 

An expression for the relation between the unit spin vector ($\hat{\bf{J}}$) and the unit traceless tidal field ($\hat{\bf{T}}$) was proposed in \citet{2000ApJ...532L...5L, 2001astro.ph.11186L}:
\beq \label{correlationparameter}
\langle\hat{J}_I\hat{J}_j|{\bf T}\rangle\equiv\frac{1+c}{3}\delta_{ij}-c\hat{T}_{ik}\hat{T}_{kj},
\eeq
where $c\in[0,3/5]$ is the correlation parameter to measure the strength of the intrinsic spin-shear alignment with the nonlinear modifications taken into account. When c=0 it corresponds to the case when nonlinear effects have completely broken down initial spin-shear correlations and when c=3/5 it is the ideal case when $\bf{I}$ is independent of $\bf{T}$.

\citet{2005ApJ...629L...5L} derived an expression using Equation \ref{correlationparameter} for the Probability Density Function (PDF) of the orientations of the galaxy spin vectors relative to the tidal spin tensors: 
\begin{equation}\begin{split} \label{pcos}
&P(\cos \alpha, \cos\beta,\cos\theta)=\frac{1}{2\pi}\prod_{i=1}^3(1+c-3c\hat{\lambda}_i^2)^{-1/2}\times\\
&\left[\frac{\cos^2\alpha}{1+c-3c\hat{\lambda}_1^2}+\frac{\cos^2\beta}{1+c-3c\hat{\lambda}_2^2}+\frac{\cos^2\theta}{1+c-3c\hat{\lambda}_3^2}\right]^{-3/2}.
\end{split}\end{equation}
Where $\hat{\lambda}_i$ are the eigenvalues of $\hat{T}$ and $\alpha$, $\beta$ and $\theta$ are the angles between the unit spin vector and the major, intermediate and minor axis of the tidal field, respectively. 

To quantify the preferred alignment of halo spins orthogonal to filament axis, we calculate $P(\cos\theta)$ which is the PDF of the cosine of the angle between spin axis and the minor axis of the tidal field which defines the axis of filaments. Filament regions are defined as having two positive and one negative eigenvector. They also must satisfy the traceless condition of $\sum_i\hat{\lambda}_i=0$ as well as the unit magnitude condition of $\sum_i\hat{\lambda}_i^2=1$. Therefore the eigenvalues in filament regions can be approximated by $\hat{\lambda}_1=\hat{\lambda}_2=1/\sqrt{6}$ and $\hat{\lambda}_3=-2/\sqrt{6}$. Using these values in Equation \ref{pcos} gives
\beq \label{PDF}
P(\cos\theta)=(1-c)\sqrt{1+\frac{c}{2}}\left[1-c\left(1-\frac{3}{2}\cos^2\theta\right)\right]^{-3/2}.
\eeq
If halo spins are oriented completely randomly then $c=0$ and the PDF is flat. If halo spins are preferentially orthogonal to filaments then $c>0$ and the function increases with $\cos\theta$. Although tidal torque theory restricts $c$ to positive values, other effects could be in play that cause halo spins to be aligned parallel with filaments, which would cause a negative value of $c$.


\begin{figure}
\plotone{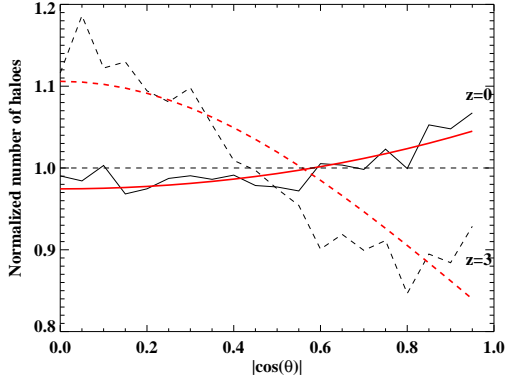}
\caption{The distribution of alignments of the spin with the axis of filaments of all dark matter haloes. The solid lines are for z=0 (these are all the haloes in the red region in Figure \ref{AMCorrelation}) and the dashed lines are for z=3 (the haloes in the green region in Figure \ref{AMCorrelation}).  For z=0, the overabundance of haloes with high values of $|\cos\theta|$ (where $\theta$ is the angle between halo spin and filament axis) indicates that haloes are preferentially aligned parallel to filaments. For z=3 there is an alignment of spins orthogonal to the axis of filaments. The red lines are fits from theory (the PDF in Equation \ref{PDF} where $c$ is the best fit value, $c=-0.035\pm0.004$ for z=0 and $c=0.129\pm0.009$ for z=3). The flat dashed line is the expected distribution for random halo spin orientations. \label{AMFilDistribution} }
\end{figure}

\subsection{Alignment of halo spin and velocity with filaments \label{filamentalignment}}

%

The alignment between a filament and the spin of the haloes that make it up is simply given by the cosine of the angle $\theta$ between the two vectors and the absolute magnitude is taken because the filament is only defined by an axis, not a particular direction. The distribution of $|\cos\theta|$ for all haloes in filaments at redshift 0 and 3 is shown in Figure \ref{AMFilDistribution} where the number of haloes in each bin of $|\cos\theta|$ is normalized to make the area under the graph unity. The shape of this distribution can be quantified in two ways; the median value or by fitting a function to the curve.

Since the distributions shown in Figure \ref{AMFilDistribution} are clearly non-Gaussian, the median rather than the mean would be the more useful statistic (although the mean was used by eg. \citet{2009ApJ...706..747Z, 2007ApJ...655L...5A}). The standard error of the median was found by bootstrap resampling and finding the standard deviation of the resampled medians. The distributions can also be fitted to the probability density function of Equation \ref{PDF} to find the correlation parameter $c$ of the intrinsic spin-shear alignment which characterizes the shape of the distribution. The fit was done using a Markov chain Monte Carlo and two examples of such a fit is shown as the red lines in Figure \ref{AMFilDistribution}.

\begin{figure}
\plotone{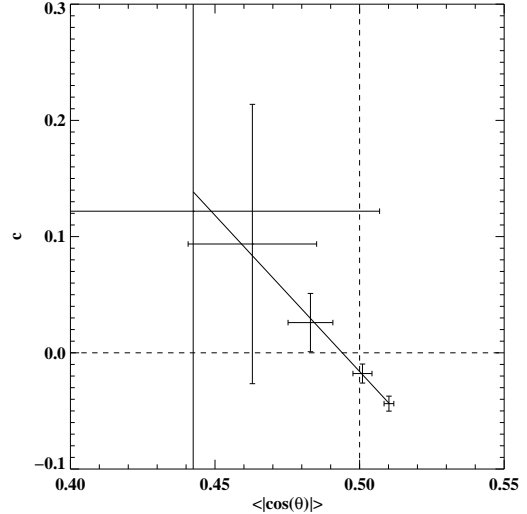}
\caption{A comparison of two ways to quantify the degree of alignment of halo spin with the large scale structure. The data points are for haloes in mass bins at z=0, scale$=2.0\,Mpc$ where the same mass bins can be seen in the red line in Figure \ref{AMCorrelation} ($\log(M)=$11.63 - 12.21, 12.21 - 12.79, 12.79 -  13.37, 13.37 - 13.95, 13.95 - 14.53). The error bars of c are the $1\sigma$ errors of the MCMC fit and the error bars of $<|cos\theta|>$ are the standard error of the median. The dashed lines are for random spin orientations.   \label{compare}}
\end{figure}

\begin{figure*}   
%
%
\begin{minipage}[l]{7in}

\plotone{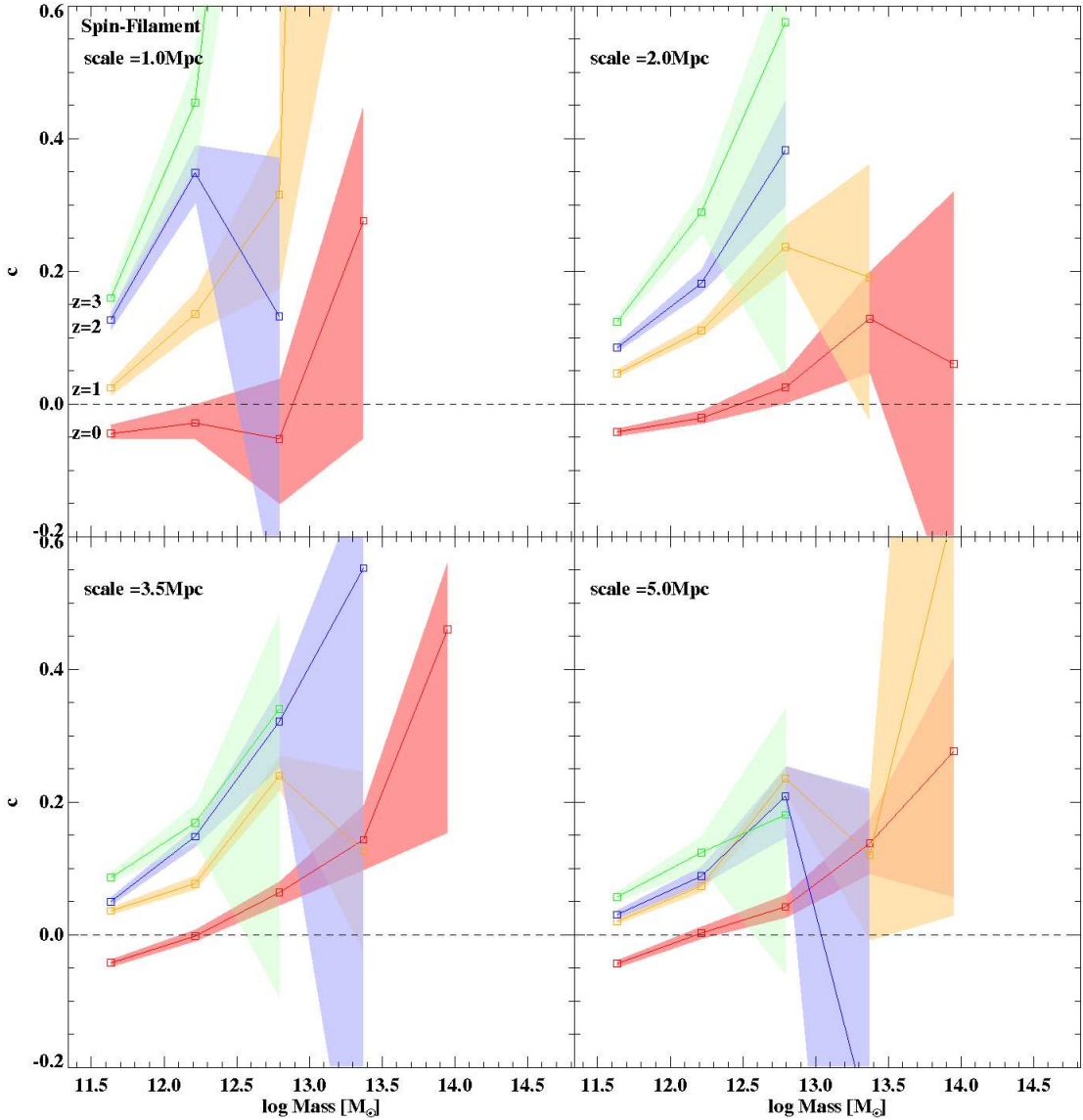}
\end{minipage}

\caption{ The alignment of dark matter halo spin with filaments over cosmic time. Alignment is characterized by the parameter $c$ of the fit of Equation \ref{PDF} to the distribution of $|\cos\theta|$, where positive $c$ indicates orthogonal alignment and negative $c$ indicates parallel alignment. The panels show filaments found in different smoothing scales: 1.0 (top left) 2.0 (top right), 3.5 (bottom left) and 5.0 Mpc/h (bottom right). At high redshift all spins are orthogonal to filaments but recent times, low mass haloes have a parallel alignment with filaments. The dashed line is the expected distribution for random halo spins and the shaded regions are the $1\sigma$ errors. The red line is for z=0, yellow line is z=1, blue is z=2 and green line is z=3. \label{AMCorrelation}}
\end{figure*}

These above two methods are compared in Figure \ref{compare} for some example points (haloes in mass bins at z=0, scale$=2.0\,Mpc$ which are the same mass bins as the red line in Figure \ref{AMCorrelation}). There is a one-to-one correlation of the two parameters so either could be used. We have chosen to use the correlation parameter $c$ in this paper since it is theoretically motivated by TTT.

The value of $c$ indicates the strength of the alignment of halo spins with the orientation of filaments, and also the intrinsic alignment of spin with the tidal field. If the haloes generally have spins parallel to filament axis $c$ is negative, conversely, if the halo spin are generally orthogonal to filament axis then $c$ will be positive. The error of $c$ is the standard deviation of the value which maximizes the likelihood of the fit of the PDF to the distribution. From the value of $c$ found for all the haloes at z=0 ($c=-0.035\pm0.004$) and for z=3 ($c=0.129\pm0.009$), the general trend is that haloes are aligned orthogonal to filaments at high redshift and aligned parallel at low redshift.

\begin{figure}
\plotone{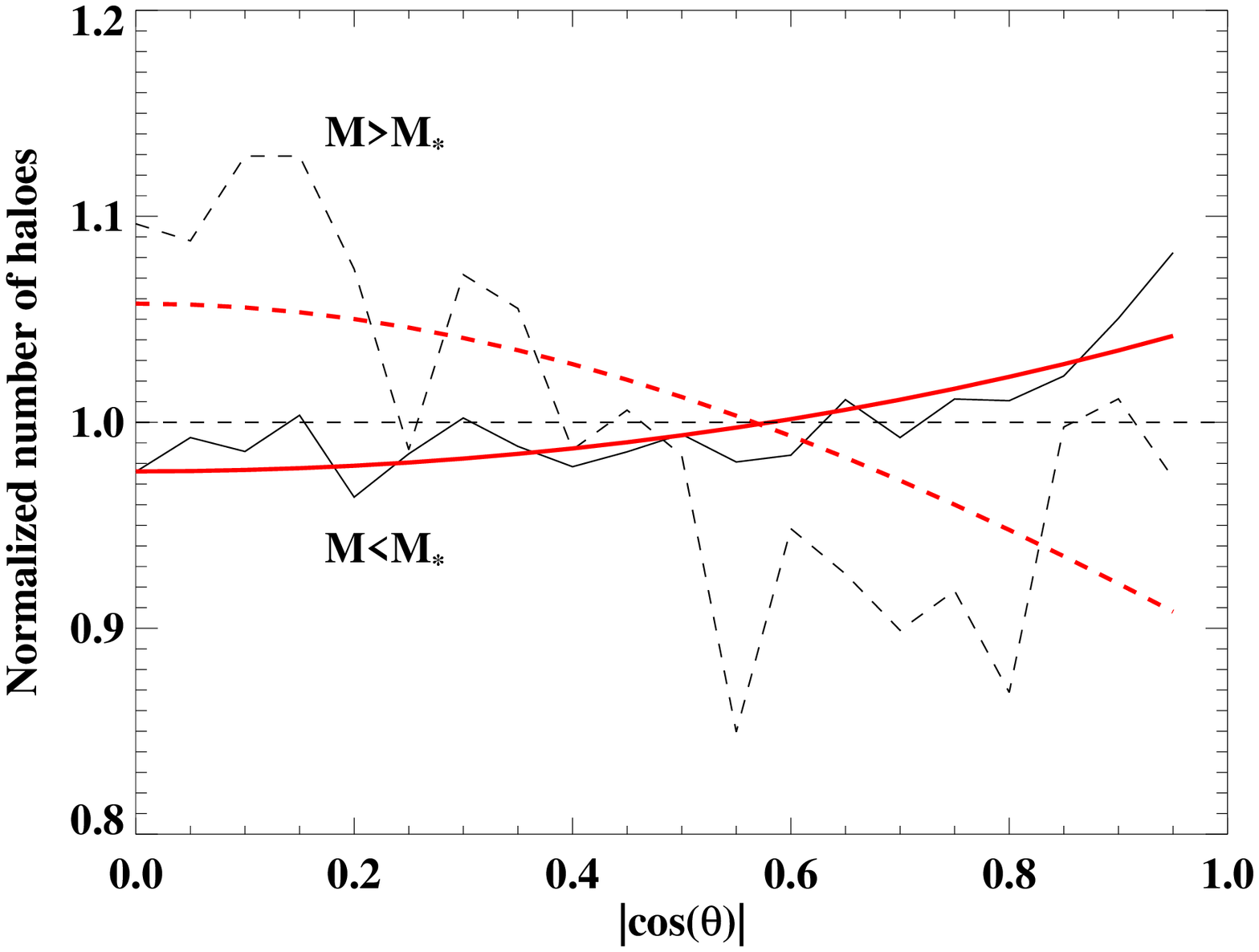}
\caption{The distribution of alignments of halo spin with the axis of filaments for low mass ($M<M_*$, solid lines) and high mass ($M>M_*$, dashed lines) haloes.  These haloes are at z=0 and filaments are found using smoothing scale 3.5Mpc/h. This mass division corresponds to the red line in the bottom left panel of Figure \ref{AMCorrelation} where the first two points of that figure are the low mass haloes and the other points are the high mass haloes.  \label{PcosthetaLowHighMass} }
\end{figure}

The alignment of halo spin vectors with filaments is shown in Figure \ref{AMCorrelation}. The alignment distribution has been fitted to find $c$ for haloes in bins of mass and for haloes at different redshifts. For all smoothing scales, it can be seen that at z=0 the alignment is weakly parallel (negative c) for low mass halos in filaments (mass less than about $M_\ast=5.89\times 10^{12} M_\odot$) and orthogonal (positive c) for high mass haloes. This is illustrated in Figure \ref{PcosthetaLowHighMass}. At higher redshifts the alignment becomes more orthogonal for all halo masses. There are less haloes in the high mass bins at high redshift because the high mass haloes have not had time to form yet. The result of \citet{2002A&A...395....1F, 2007ApJ...655L...5A, 2007MNRAS.375..489H} and \citet{ 2009ApJ...706..747Z} that halo spins generally lie along the axis of filaments is driven by the low mass haloes at z=0. This is demonstrated in Figure \ref{AMFilDistribution} where the alignment distribution for all haloes at z=0 is shown. The alignment is preferentially parallel because of the high number of low mass haloes that exhibit parallel alignment.

The affects of smoothing scale on the halo spin alignment with filaments show something about the formation of filaments. For redshift 0 (the red line in Figure \ref{AMCorrelation}), haloes seem to be best aligned at a large smoothing scale while high redshift haloes are best aligned at small scales. If an orthogonal alignment is an indicator that a halo formed inside a filament topology, then this shows that filaments grow in size over time.

\begin{figure}
\plotone{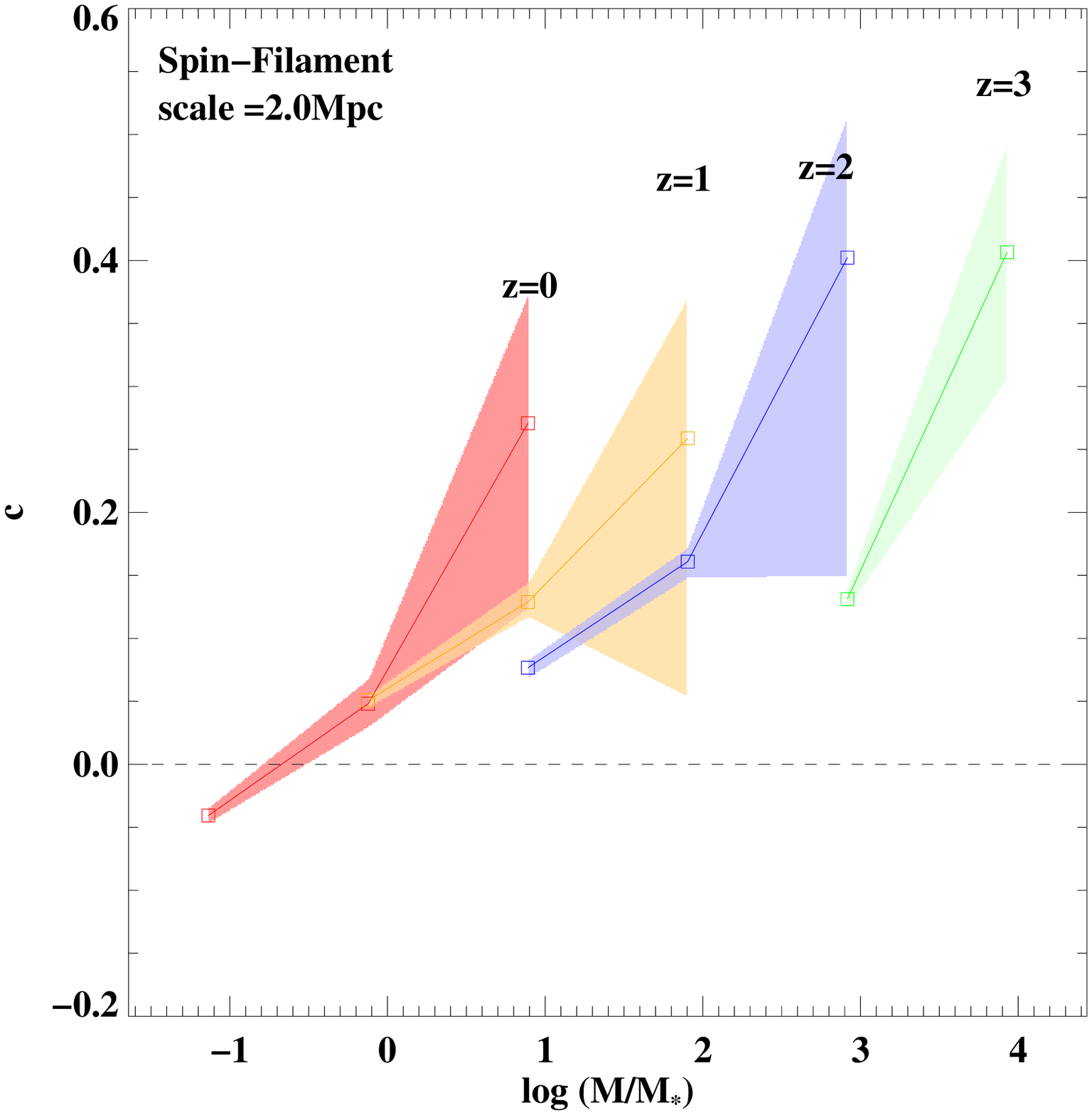}
\caption{The alignment of dark matter halo spin with filaments over cosmic time for haloes in bins of halo mass normalized by the characteristic mass. The alignment for a 2Mpc scale is shown here.
\label{AMCorrelationMchar}}
\end{figure}

Figure \ref{AMCorrelationMchar} shows the effect of taking into account the characteristic mass. Here we can compare haloes between redshifts at equivalent stages of collapse. When the this is accounted for, almost all the points overlap within their errors. This means that haloes at a similar stage in their collapse have the same degree of preferential alignment with filaments over cosmic time. A halo that is just starting to collapse ($M=M_\ast$) at redshift 2 has a similar probability of orthogonal alignment with its filament as a halo that is just starting to collapse at redshift 1 or 0. However, no assumptions were made about the evolving scale of filaments and the smoothing scale was kept constant at 2.0Mpc. Even with a constant scale, this similarity between alignments at different times shows that the build up of spin is closely linked with a halo's formation.

When substructure is discounted by taking the most massive subhalo in each FOF group, there is practically no change in the alignments.

\begin{figure*}   

\begin{minipage}[l]{7in}

\plotone{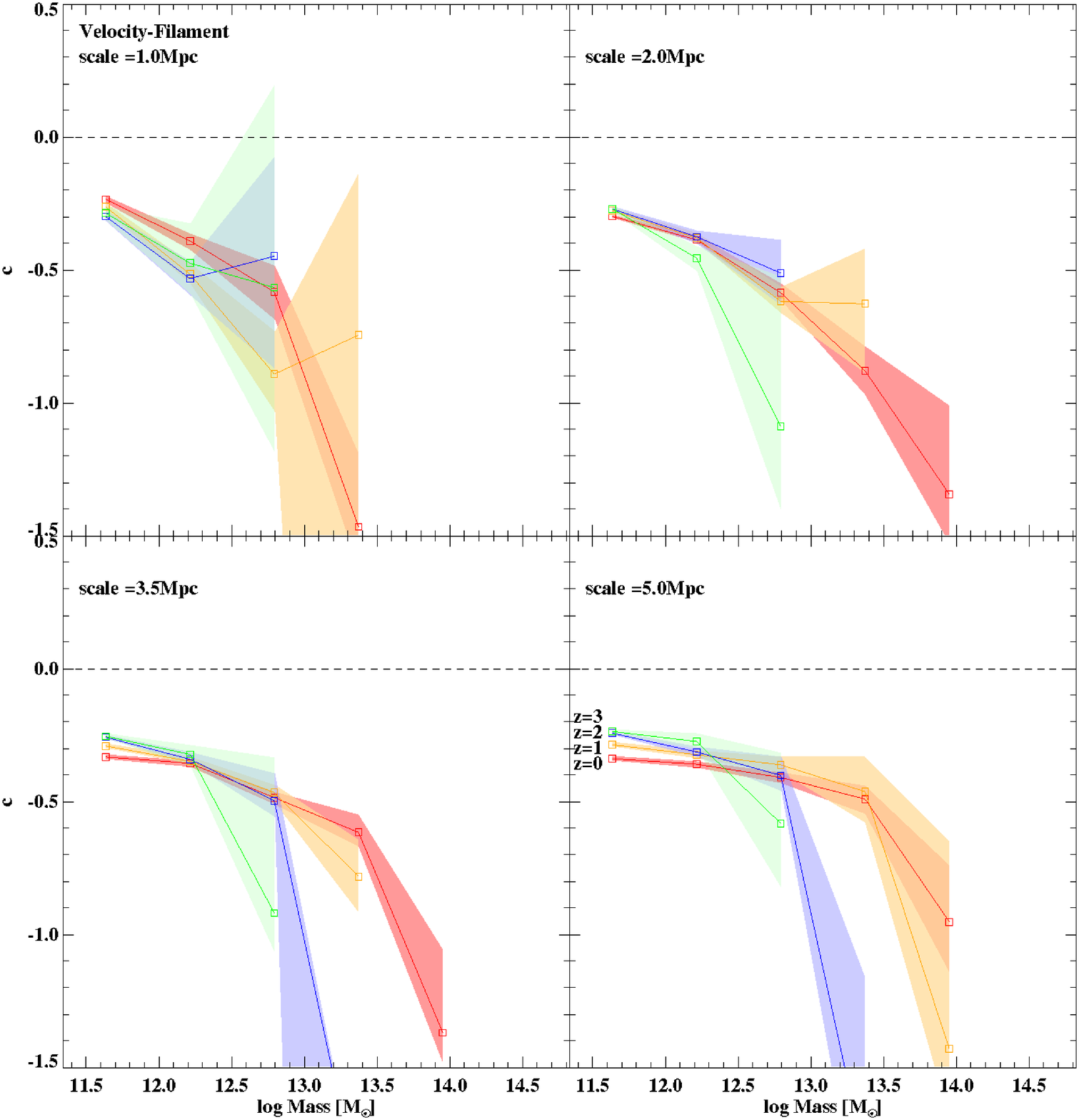}
\end{minipage}

\caption{ The alignment of dark matter halo velocity with filaments. For all redshifts, haloes are parallel aligned with filaments which demonstrates a streaming motion of haloes down bulk flows. Alignment is characterized by the $c$ parameter of Equation \ref{PDF} where $\theta$ is the angle between halo velocity and filament axis.  Lines are colored as in Figure \ref{AMCorrelation}. \label{VelocityCorrelation}}
\end{figure*}

\begin{figure}
\plotone{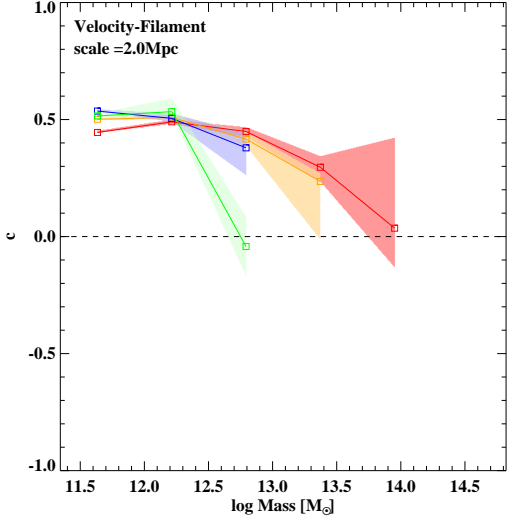}
\caption{ The alignment of dark matter halo velocity with filaments on the scale of 2.0Mpc where bulk motions have been subtracted.  Colored lines are for different redshifts as in Figure \ref{AMCorrelation}. \label{VelocityCorrelationBULKMOTION}}
\end{figure}

\begin{figure}
\plotone{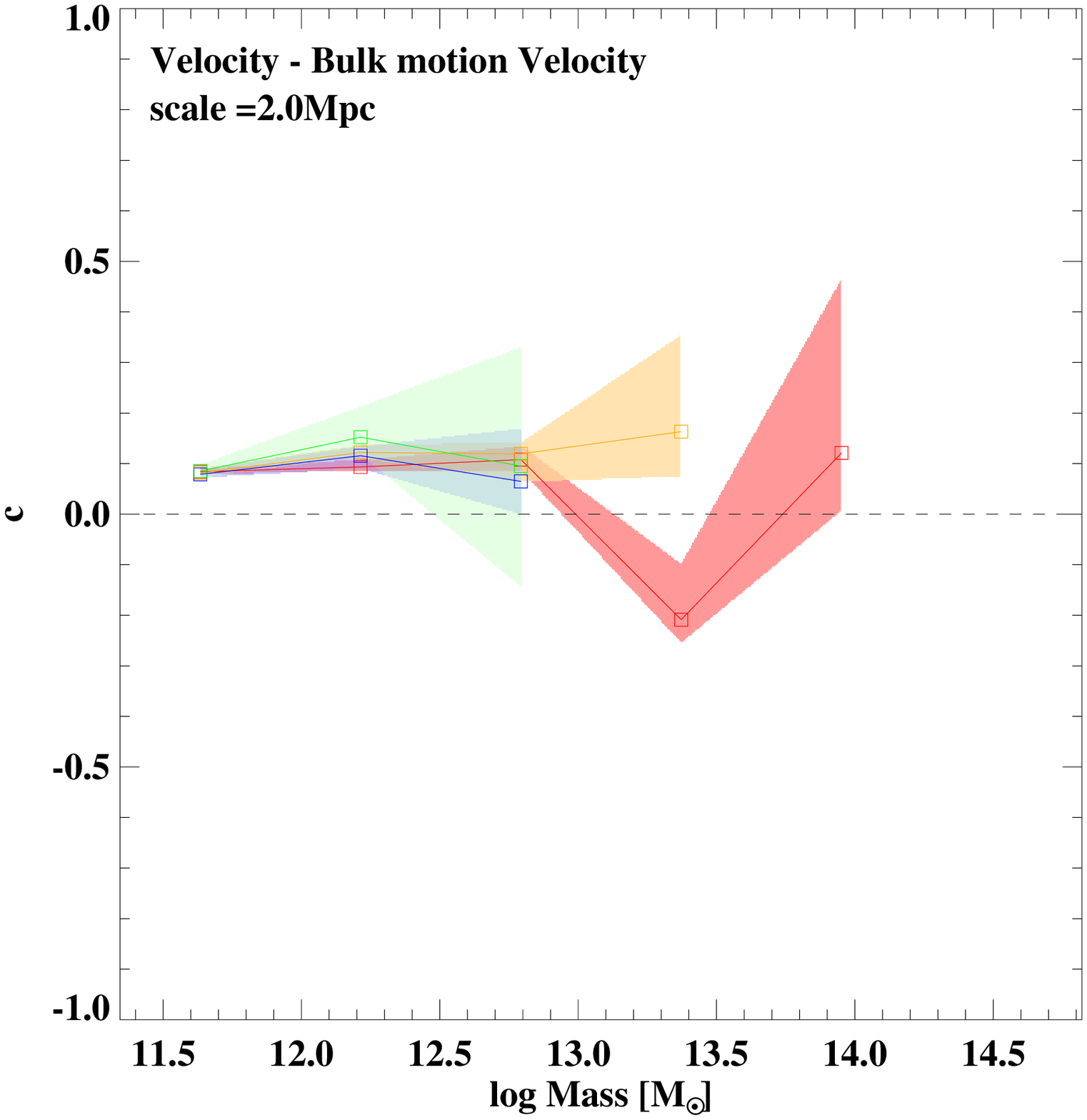}
\caption{ The alignment of dark matter halo velocity with the local bulk motion on the scale of 2.0Mpc.  Colored lines are for different redshifts as in Figure \ref{AMCorrelation}. \label{BULKMOTION}}
\end{figure}

Although the $c$ parameter was introduced in the context of spin alignments with the tidal field (manifested by filaments in the large scale structure), it can also be used as a more general measure of alignment. The distributions of $|\cos\theta|$ where $\theta$ is the angle between halo centre of mass velocity and filament axis is also well-fit by the PDF in Equation \ref{PDF}. Again, a negative value of $c$ means a parallel alignment and a positive value is orthogonal alignment.

All panels of Figure \ref{VelocityCorrelation} show a parallel alignment which is stronger for high mass haloes. This shows streaming of haloes of all masses down filaments into massive clusters.

This streaming can be seen in the velocity vectors of haloes in some filaments in the middle panel of Figure \ref{vectors}, where vectors are pointed along filaments towards clusters. However, some filaments display bulk motions where the entire filament is moving towards some attractor. To see the extent of these bulk motions, they have been subtracted in Figure \ref{VelocityCorrelationBULKMOTION} by subtracting the mass-weighted average velocity of haloes by halo mass found within the smoothing scale on which the filaments were found. When bulk motions are discarded, an orthogonal motion remains. The apparent streaming of haloes down filaments was wholly caused by bulk motions of entire filaments, and this bulk flow is generally along the axis of filaments. The relative motions can be seen in Figure \ref{BULKMOTION} in the alignment of halo velocity with the flow of the local bulk motion. (Bulk motions have been subtracted from halo velocities here.) Low mass haloes are moving slightly orthogonal to the flow and high mass haloes have no preferred direction of motion. This reflects how bulk motions have been removed: high mass haloes were given more weight than low mass haloes and so the residual motions of high mass haloes once bulk flow is removed is minimal.


%
 The enlargement of filaments over time that was seen in the spin alignments is also visible in the way the bulk flows are aligned. The low mass haloes at z=0 (red line in Figure \ref{VelocityCorrelation}) are more strongly aligned at large smoothing scales and the low mass haloes at high redshifts are most aligned at small smoothing scales. If filaments are chutes where haloes are channeled into clusters then these low mass haloes are evidence for the growth of the size of filaments over time. The high mass haloes on the other hand are generally less aligned at large smoothing scales for all redshifts which is seen as a flattening of the curves. This may be due to the inclusion of some cluster haloes when the smoothing scale is broadened which would introduce random velocities into the sample.
%


Although both halo spin and velocity are somewhat aligned with filaments, these alignments are not strong enough so that there is a significant alignment between a halo's spin and velocity.


\subsection{Halo-halo spin alignment }\label{halohaloalignment}

Tidal torque theory predicts that as well as being aligned with the large scale structure, halo spins should be aligned with each other. This is usually tested by simply taking the average of the dot product of pairs of halo spins separated by distance r;

\begin{equation} \label{eta}
	\eta(r)=\langle |\hat{\bf{J}}(x)\cdot\hat{\bf{J}}(x+r)|\rangle.
\end{equation}

A second quantity used by \citet{2000ApJ...543L.107P} and \citet{2005ApJ...627..647B} is 

\begin{equation}\label{eta2}
	\eta_2(r)=\langle |\hat{\bf{J}}(x)\cdot\hat{\bf{J}}(x+r)|^2\rangle -\frac{1}{3}.
\end{equation}

These quantities are plotted in the top panels of Figure \ref{AMAM}, where at very small halo separations ($r<0.3$Mpc/h) there seems to be a parallel alignment of halo spins. 

\begin{figure}
\plotone{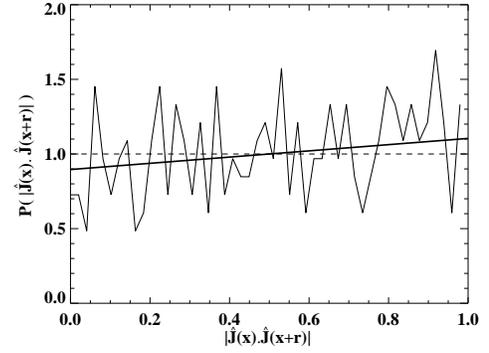}
\caption{An example of a distribution of halo-halo spin alignments. $P\left(|\hat{\bf{J}}(x)\cdot\hat{\bf{J}}(x+r)|\right)$ is the number of halo pairs in each bin of $\hat{\bf{J}}\cdot\hat{\bf{J}}$. This example is for haloes that are separated from 0.06 to 0.1 Mpc/h, which is the second data point from the left in Figure \ref{AMAM}. The thin line is the actual distribution and the thick line is a straight line fit. There is a significant deviation from random spin orientations here, shown by the positive slope of the straight line. \label{AMAMDistribution} }
\end{figure}

\begin{figure}
\plotone{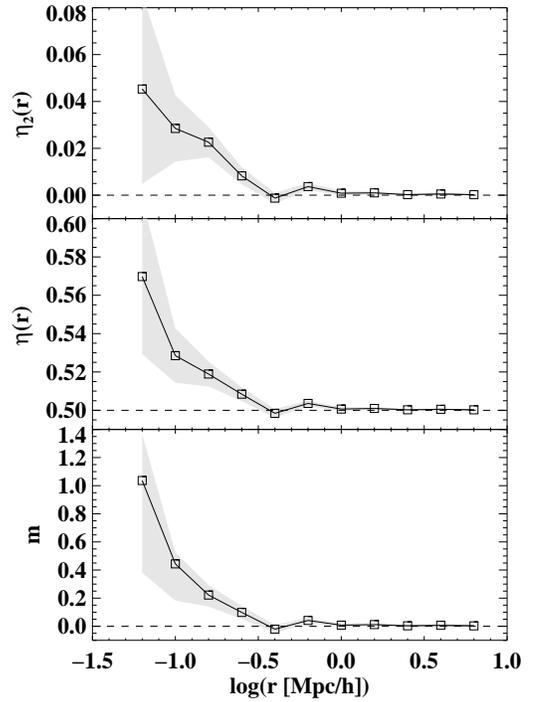}
\caption{The alignment of neighbouring halo's spins, for haloes separated by distance r. Three different statistics are used: $\eta$ from Equation \ref{eta} (top), $\eta_2$ from Equation \ref{eta2} (middle), and $m$, the slope of the distribution of the halo-halo spin alignment (bottom, an example is shown in Figure \ref{AMAMDistribution}). The dashed lines are for random halo alignments and the shaded regions are the $1\sigma$ errors.
 \label{AMAM} }
\end{figure}

\begin{figure}
\plotone{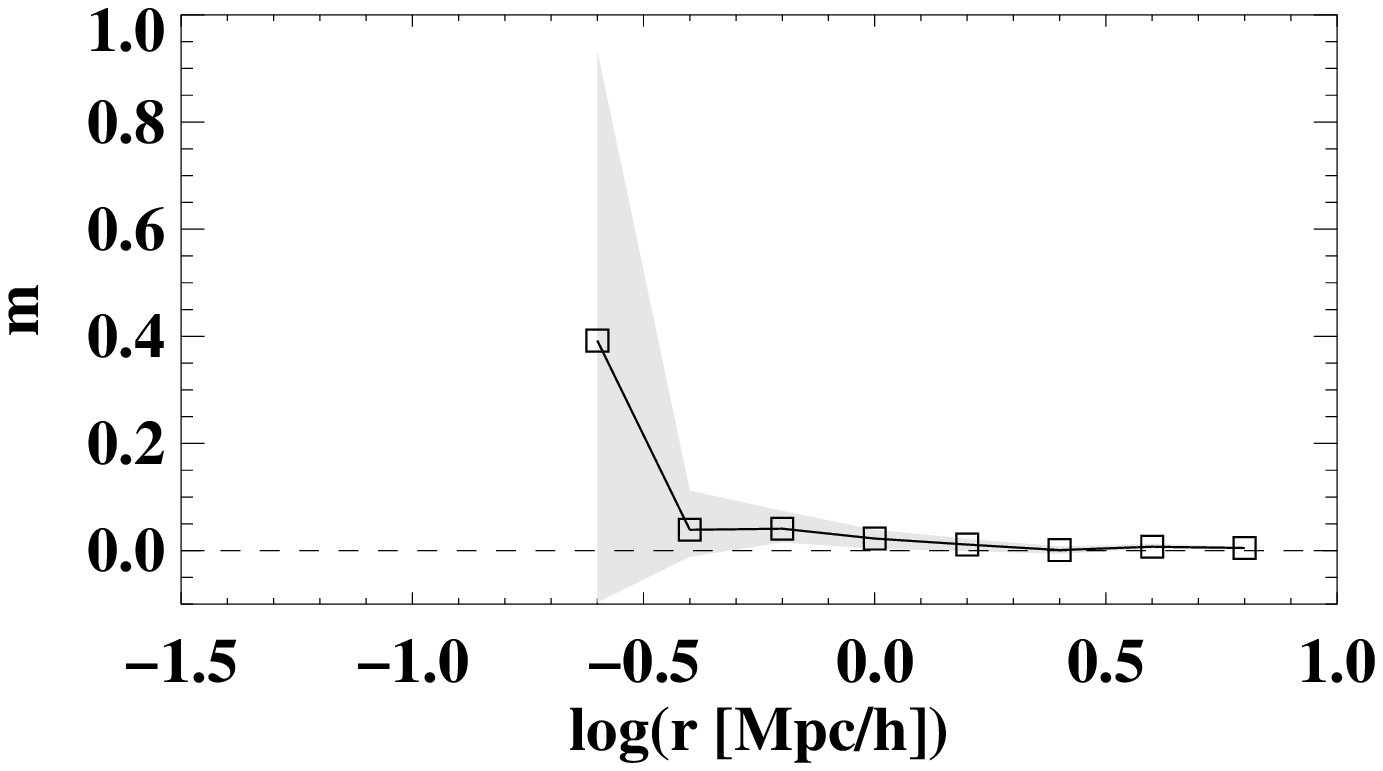}
\caption{The alignment of neighbouring friends-of-friends halo's spins. There is no alignment at any scale.
 \label{AMAMFOF} }
\end{figure}

However, both of these quantities rely on taking an average over all the halo pairs in each bin of separation. The mean is a useful value when dealing with a peaked distribution, but none of the actual distributions of $|\hat{\bf{J}}(x)\cdot\hat{\bf{J}}(x+r)|$ has an apparent peak (an example of one of these distributions is Figure \ref{AMAMDistribution}, where $P(\hat{\bf{J}}\cdot\hat{\bf{J}})$ is the number of haloes in each bin normalized so that the area under the curve is unity). A fairer way of dealing with these noisy distributions is to fit a straight line and see if there is any deviation from randomness. The slope of the best fit line indicates if more haloes are aligned parallel or orthogonal to each other. 
\begin{equation}
 P(|\hat{\bf{J}}(x)\cdot\hat{\bf{J}}(x+r)|)=m|\hat{\bf{J}}(x)\cdot\hat{\bf{J}}(x+r)|+c.
\end{equation}
A positive slope ($m$) of the best fit line means there are more parallel aligned halo pairs, a negative $m$ means they are more orthogonal and $m=0$ means the haloes have random alignment. The values of $m$ that maximized the likelihood of fitting a straight line to the distributions are shown in the bottom panel of Figure \ref{AMAM}.

The shape of the plot of the slope (bottom panel of Figure \ref{AMAM}) is similar to the shape of the plots of the conventional statistics. This is expected since they are effectively measuring the same thing but in a slightly different way. Halo spins are aligned parallel for halo separations under $0.3$Mpc/h. This alignment has not been seen before in simulations because it exists only on very small scales which have not before been examined. It has however been seen in galaxy surveys, for example Galaxy Zoo \citep{2009MNRAS.392.1225S} found alignment for galaxies closer than $0.5$Mpc. The alignment exists on the scale of substructure within clusters. If only the most massive subhalo in each FOF group is taken (the substructure is thrown out), then there is no significant alignment at any scale (Figure \ref{AMAMFOF}). Here there are no haloes at small separations and there is no significant alignment at any scale. Only the subhaloes within large clusters exhibit any halo-halo spin alignment, although it is weak.

\section{Evolution of spin parameter }\label{SPSection}

\begin{figure}
\plotone{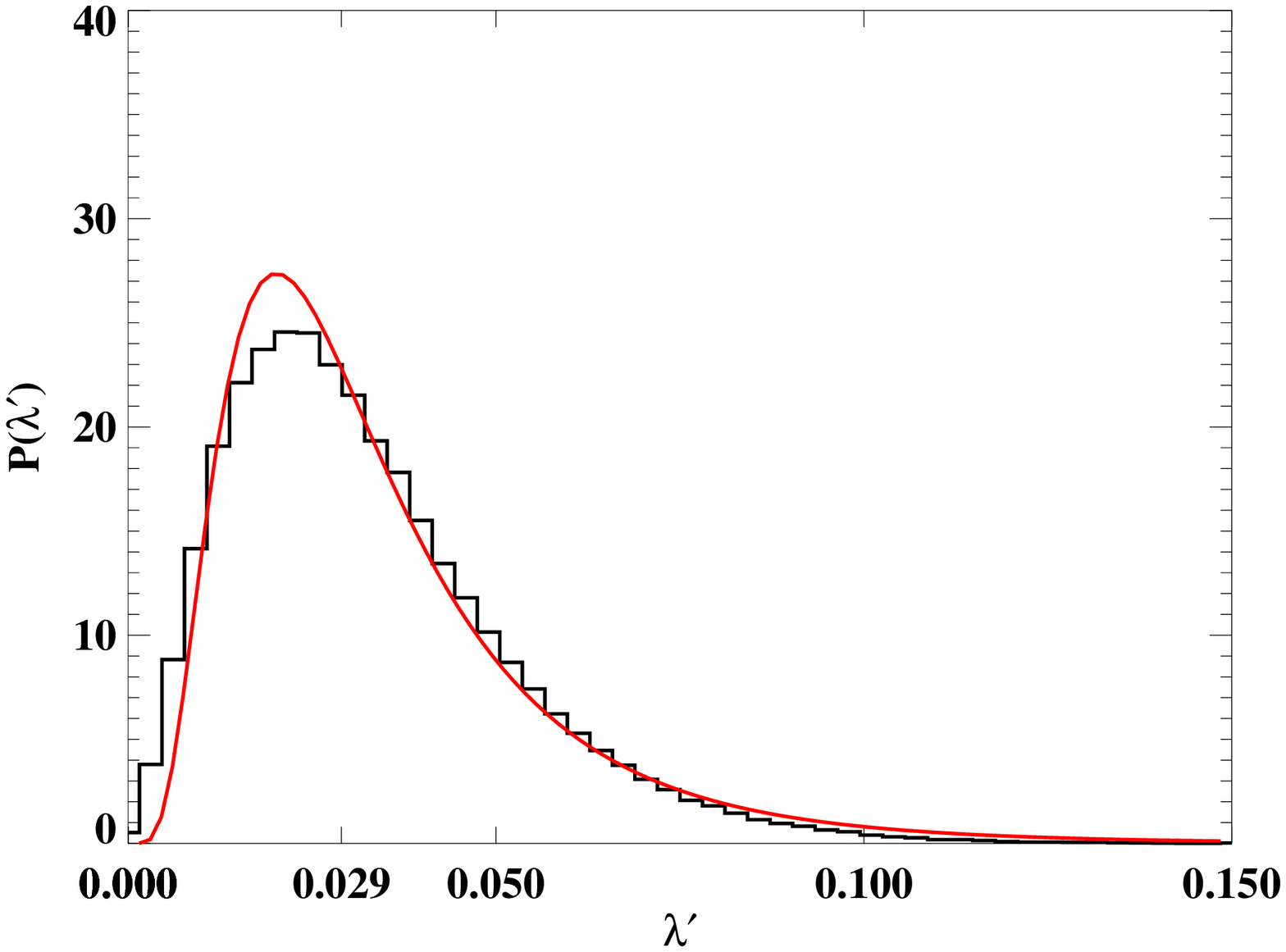}
\caption{The distribution of the spin parameter at z=0. The histogram is the data and the red (smooth) line is a log-normal fit (Equation \ref{lognormal}) where the best fit values are $\lt_{0} =0.02900^{+0.00006}_{-0.00005}$,  $\sigma= 0.604 ^{+0.001}_{-0.002}$.
. \label{SpinParameter} }
\end{figure}

\begin{figure*}
\begin{minipage}[l]{7in}
\plottwo{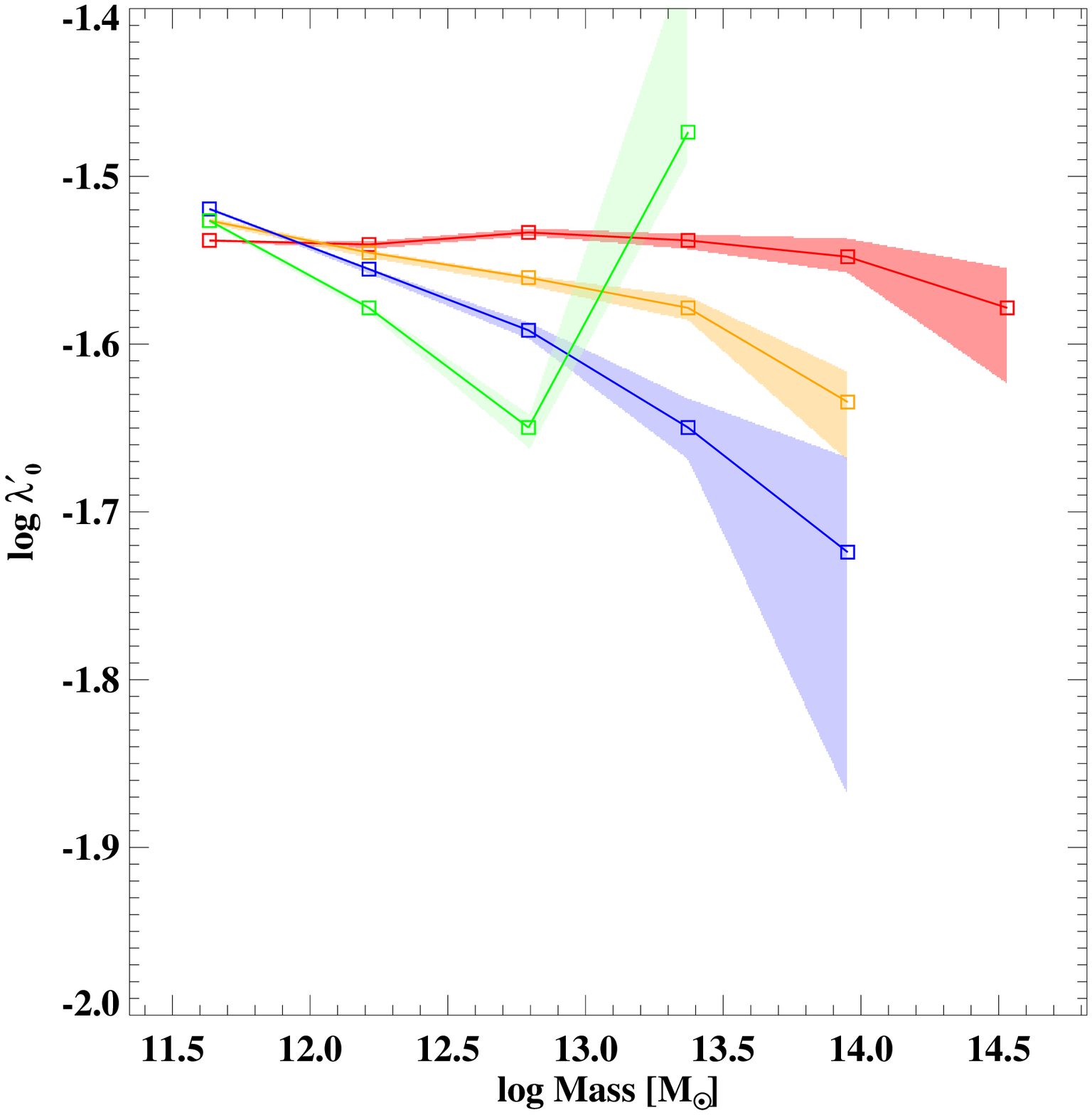}{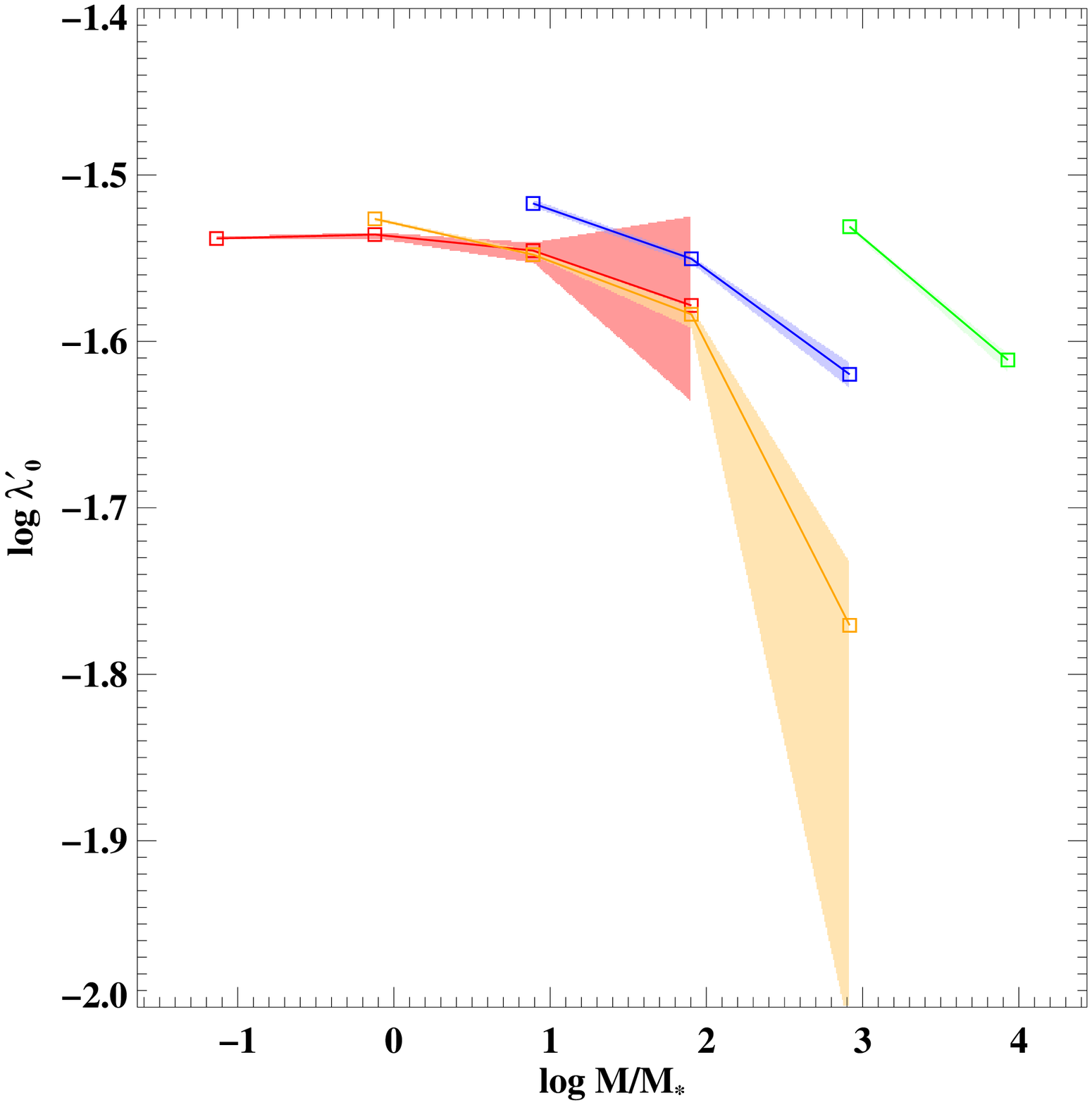}
\end{minipage}
\caption{The redshift evolution of the spin parameter ($\lt_{0}$). The red line is for z=0, yellow is z=1, blue is z=2 and green is z=3. The shaded regions are the $1\sigma$ confidence intervals. \textbf{Left}: At high redshift the spin parameter is less and there is a mass dependency. \textbf{Right}: Mass bins are normalized by the characteristic mass. 
 \label{SpinParameterEvolution} }
\end{figure*}

The spin parameter is a measure of the amount of angular momentum contained in a halo. It was defined in \citet{2001ApJ...555..240B} as,
\begin{equation}
	\lt \equiv \frac{|\bf{J}|}{\sqrt{2}MVR}
\end{equation}
given the angular momentum $\bf{J}$ inside a sphere of radius $R$ 
containing mass $M$, and where $V$ is the halo circular velocity at radius $R$, $V^2=GM/R$.

The distribution of $\lt$ over the halos in our sample
is shown in Figure \ref{SpinParameter}. It is well fit by a log-normal
distribution, 
\begin{equation}\label{lognormal}
        P(\lt) = \frac{1}{\lt\sqrt{2\pi}\sigma}\exp\left(-\frac
        {\ln^2(\lt/\lt_{0})}{2\sigma^2}\right).
\end{equation}
The fit was done using a Markov chain Monte Carlo maximum likelihood analysis. For all haloes with more than 500 particles at z=0 the best fit values are $\lt_{0} = 0.02900^{+0.00006}_{-0.00005}$,  $\sigma= 0.604 ^{+0.001}_{-0.002}$ and  at z=3 $\lt_{0} = 0.02940^{+0.00008}_{-0.0001}$, $\sigma=0.576\pm0.002$. The distributions at both these redshifts over all haloes in the snapshots are nearly identical.

When haloes are binned by mass, the spin parameter at high redshift shows a mass dependence while there is no mass dependence at z=0, as shown in the left hand side of Figure \ref{SpinParameterEvolution}. Here the spin parameter is characterized by the mid point of the log-normal distribution, $\lt_{0}$. The spin parameter over all redshifts is only the same for low mass ($M<10^{12}$) haloes but there are far more low mass than high mass haloes. Since low mass haloes dominate, the average distributions over all haloes at the different redshifts look the same. At high redshift, there is a tendency for the spin parameter to be smaller for high mass haloes.

This redshift dependency can be characterized by a power relationship between $\lt_{0}$ and mass at each redshift;
\begin{equation}
\lt_{0}\propto M^{a(z)}.
\end{equation}

\begin{figure}
\plotone{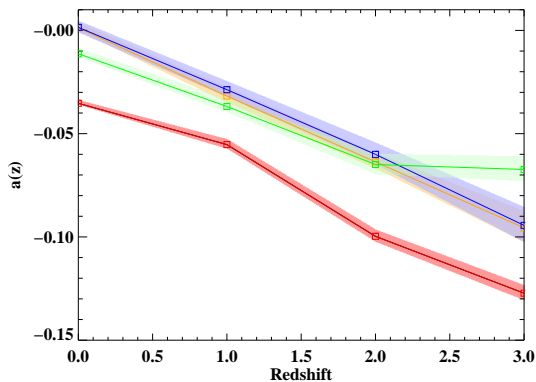}
\caption{The mass dependence of the spin parameter over redshift. The mass dependence, $a$ is the slope of the straight dashed lines in Figure \ref{SpinParameterEvolution}. The red (lowest) line includes all haloes with more than 100 particles, orange line includes 500 particles and blue line includes 1000 particles. The green line is for the halo catalogue which doesn't include substructure. There is a linear trend of stronger mass dependence at higher redshift. \label{massdep} }
\end{figure}

The more negative the value of $a$, the stronger the correlation and $a=0$ is no correlation at all. The redshift dependence of $a$ is shown in  Figure \ref{massdep}. The lines for haloes with $>500$ particles and $>1000$ particles overlap in Figure \ref{massdep} whereas the line for haloes with $>100$ particles does not. This shows that haloes with more than 100 particles are susceptible to errors from particles in the outer regions and the cut off of only using haloes with more than 500 particles is justified. 

\citet{2008ApJ...678..621K} found that mass binning and selection criteria for relaxed haloes has almost no effect on this correlation. We did find a small effect when a different halo catalogue was used. Instead of using all the subhaloes, only the most massive subhalo (with more than 500 particles) in each friends-of-friends halo was used. Most of the mass of the FOF halo is in the most massive subhalo so it can be regarded as the background halo itself. When substructure is disregarded, we find that there is a stronger mass dependency of the spin parameter at almost all redshifts (the green line in Figure \ref{massdep} is below the corresponding orange line which includes all substructure). The spins of subhaloes are greatly affected by interactions and merger events so may be out of equilibrium.

Mass dependence of the spin parameter at high redshift was first found by \citet{2008ApJ...678..621K}, who looked at z=1 and z=10. When extrapolating the linear trend of $a(z)$ with redshift, we predict a much stronger correlation, $a(z=10)\simeq-3$ whereas they find  $a(z=10)=-0.059\pm0.171$. Our results agree more closely with  \citet{2011MNRAS.411..584M} who found $a(z=2)\simeq-0.03$. For haloes in different environments (blobs, filaments, sheets and voids), the trends are the same.

When halo mass is scaled by characteristic mass in the right hand side of Figure \ref{SpinParameterEvolution}, we find that haloes at similar stages of collapse at z=0 and 1 have the same spin parameter (the orange and red lines overlap). At high redshift, haloes at similar stages of collapse have a higher spin parameter (At $\log M/M_\ast=3$ for example, the green (z=3) point lies above the points for z=2 and z=1). This may be the result of accretion and merger events decreasing the spin of haloes. At z=3, haloes have retained much of their initial spin but by z=1, similar haloes have experienced accretion that has lowered their spin parameter.


\section{Summary and Discussion}\label{summary} 

Using the Millennium N-body simulation, we have tracked the evolution of dark matter halo angular momentum alignments with the large scale structure, with each other and the evolution of the spin parameter. We have used the shape of the density field to find filaments of ~2Mpc in scale in the large scale structure. The alignment between dark matter halo spin and the axis of filaments was characterized by the shape of the distribution of $|cos(\theta)|$ where $\theta$ is the angle between the two vectors. The distribution was fitted to the PDF of Equation \ref{PDF} to find the free parameter $c$ which characterized the strength of parallel or orthogonal alignment.

We found that angular momentum vectors of dark matter haloes since z=3 are generally orthogonal to filaments but high mass haloes have a stronger orthogonal alignment than low mass haloes. At z=0 the spins of low mass haloes have become parallel to filaments, whereas high mass haloes keep their orthogonal alignment.

An interpretation of this is that at early times all halo spins were aligned orthogonal to filaments, as TTT predicts. High mass haloes especially are well aligned because they have had their maximal expansion more recently and so will have been tidally torqued for longer. They usually exist close to clusters where the infall of dark matter is almost isotropic and so the nett effect from mergers and accretion is minimal. Low mass haloes, however, are vulnerable to being disturbed by mergers and accretion which is usually assumed to have the effect of randomizing the spin orientation. This leaves unexplained why low mass haloes at low redshift exhibit a parallel alignment with filaments.

We found that filaments are regions bulk flow. When bulk flows are included there is a clear trend for haloes to travel parallel to filaments, and high mass haloes travel with the best alignment.  When bulk flows on the scale of the filaments are subtraced, an orthogonal alignment to filaments remains, particularly for low mass haloes. This shows that entire filaments themselves are moving towards attractors and on small scales there is only orthogonal motion. There was also an orthogonal motion of low mass haloes with the bulk flow but no alignment of high mass haloes out of the bulk flow.

The motions of haloes relative to the bulk flow could affect how matter is accreted onto them and the spin orientation this would cause. Orthogonal motion to the bulk flow and filaments by low mass haloes could cause low mass haloes to accrete matter preferentially in one direction. High mass haloes traveling with the bulk flow would experience accretion differently, and this could cause the difference in spin orientation.

Filaments at large smoothing lengths at low redshift contain haloes with the best aligned spins and bulk motion, while at high redshift it is filaments at small smoothing lengths that contain the best aligned haloes. This shows that filaments are growing in size over time. Because of the nature of the way that the filaments were found (using Gaussian smoothing), this enlargement tells more about the width of the filaments rather than the length. This is complimentary to \citet{2009MNRAS.393..457S} where filament length is discussed and it was found that there is a general dilation of filaments that began larger and a shrinking, fusion and disappearance of the smaller filaments.

We found an alignment only between the spin orientation of very close neighbouring haloes. Only at separations of less than $0.3$Mpc/h do haloes exhibit any mutual parallel alignment of their spin axis. The halo finding method used in the Millennium simulation has enabled us to see this small scale alignment. In the Millennium simulation, the subfind algorithm was used to identify substructure in friends-of-friends groups, and the subhaloes are counted as haloes. This means that alignments between very close haloes can be probed, not just alignments between the friends-of-friends groups.

Lastly, we tracked the evolution of the spin parameter from z=3 to now and its dependence on halo mass. This was done by finding the centre of the log-normal distribution of the spin parameter. There is a mass dependace of the spin parameter at z=3 but not at low redshift and the spin parameter is lower overall at high redshift. The spin parameter follows a power law with halo mass at high redshift but is independent of mass at z=0. 

Future work will bridge the gap between idealistic CDM simulations and real galaxy observations. To do this we will generate mock galaxy catalogues and use only the data that would be available in a real survey, see if any alignments of galaxy spin orientations could be seen in the universe. This could be used to plan a survey using new multi-object IFU instruments \citep{2011OExpr..19.2649B,Croom}.

\acknowledgments
The Millennium Simulation databases used in this paper and the web application providing online access to them were constructed as part of the activities of the German Astrophysical Virtual Observatory. We would like to thank Erwin Platen and Rien van de Weygaert for providing us with the DTFE code. HET would like to acknowledge the support of an Australian Postgraduate Award and a Denison Merit Award. GFL gratefully acknowledges the Australian Research Council for support through DP0665574 and DP110100678, and for his Future Fellowship (FT100100268). JBH is supported by a Federation Fellowship from the Australian
Research Council. We would like to thank the anonymous referee for the helpful suggestions.



\begin{thebibliography}{50}
\expandafter\ifx\csname natexlab\endcsname\relax\def\natexlab#1{#1}\fi

\bibitem[{{Arag{\'o}n-Calvo} {et~al.}(2007{\natexlab{a}}){Arag{\'o}n-Calvo},
  {Jones}, {van de Weygaert}, \& {van der Hulst}}]{MMF}
{Arag{\'o}n-Calvo}, M.~A., {Jones}, B.~J.~T., {van de Weygaert}, R., \& {van
  der Hulst}, J.~M. 2007{\natexlab{a}}, \aap, 474, 315

\bibitem[{{Arag{\'o}n-Calvo} {et~al.}(2007{\natexlab{b}}){Arag{\'o}n-Calvo},
  {van de Weygaert}, {Jones}, \& {van der Hulst}}]{2007ApJ...655L...5A}
{Arag{\'o}n-Calvo}, M.~A., {van de Weygaert}, R., {Jones}, B.~J.~T., \& {van
  der Hulst}, J.~M. 2007{\natexlab{b}}, \apjl, 655, L5

\bibitem[{{Bailin} \& {Steinmetz}(2005)}]{2005ApJ...627..647B}
{Bailin}, J. \& {Steinmetz}, M. 2005, \apj, 627, 647

\bibitem[{{Bett} {et~al.}(2010){Bett}, {Eke}, {Frenk}, {Jenkins}, \&
  {Okamoto}}]{2010MNRAS.404.1137B}
{Bett}, P., {Eke}, V., {Frenk}, C.~S., {Jenkins}, A., \& {Okamoto}, T. 2010,
  \mnras, 404, 1137

\bibitem[{{Bett} \& {Frenk}(2011)}]{2011arXiv1104.0935B}
{Bett}, P.~E. \& {Frenk}, C.~S. 2011, ArXiv e-prints

\bibitem[{{Bland-Hawthorn} {et~al.}(2011){Bland-Hawthorn}, {Bryant},
  {Robertson}, {Gillingham}, {O'Byrne}, {Cecil}, {Haynes}, {Croom}, {Ellis},
  {Maack}, {Skovgaard}, \& {Noordegraaf}}]{2011OExpr..19.2649B}
{Bland-Hawthorn}, J., {Bryant}, J., {Robertson}, G., {Gillingham}, P.,
  {O'Byrne}, J., {Cecil}, G., {Haynes}, R., {Croom}, S., {Ellis}, S., {Maack},
  M., {Skovgaard}, P., \& {Noordegraaf}, D. 2011, Optics Express, 19, 2649

\bibitem[{{Brunino} {et~al.}(2007){Brunino}, {Trujillo}, {Pearce}, \&
  {Thomas}}]{2007MNRAS.375..184B}
{Brunino}, R., {Trujillo}, I., {Pearce}, F.~R., \& {Thomas}, P.~A. 2007,
  \mnras, 375, 184

\bibitem[{{Bullock} {et~al.}(2001){Bullock}, {Dekel}, {Kolatt}, {Kravtsov},
  {Klypin}, {Porciani}, \& {Primack}}]{2001ApJ...555..240B}
{Bullock}, J.~S., {Dekel}, A., {Kolatt}, T.~S., {Kravtsov}, A.~V., {Klypin},
  A.~A., {Porciani}, C., \& {Primack}, J.~R. 2001, \apj, 555, 240

\bibitem[{{Cervantes-Sodi} {et~al.}(2008){Cervantes-Sodi}, {Hernandez}, {Park},
  \& {Kim}}]{2008MNRAS.388..863C}
{Cervantes-Sodi}, B., {Hernandez}, X., {Park}, C., \& {Kim}, J. 2008, \mnras,
  388, 863

\bibitem[{{Croom} {et~al.}(2011)}]{Croom}
{Croom}, S. {et~al.} 2011, \mnras, submitted

\bibitem[{{Cuesta} {et~al.}(2008){Cuesta}, {Betancort-Rijo}, {Gottl{\"o}ber},
  {Patiri}, {Yepes}, \& {Prada}}]{2008MNRAS.385..867C}
{Cuesta}, A.~J., {Betancort-Rijo}, J.~E., {Gottl{\"o}ber}, S., {Patiri}, S.~G.,
  {Yepes}, G., \& {Prada}, F. 2008, \mnras, 385, 867

\bibitem[{{Davis} {et~al.}(1985){Davis}, {Efstathiou}, {Frenk}, \&
  {White}}]{fof}
{Davis}, M., {Efstathiou}, G., {Frenk}, C.~S., \& {White}, S.~D.~M. 1985, \apj,
  292, 371

\bibitem[{{Faltenbacher} {et~al.}(2002){Faltenbacher}, {Gottl{\"o}ber},
  {Kerscher}, \& {M{\"u}ller}}]{2002A&A...395....1F}
{Faltenbacher}, A., {Gottl{\"o}ber}, S., {Kerscher}, M., \& {M{\"u}ller}, V.
  2002, \aap, 395, 1

\bibitem[{{Gardner}(2001)}]{2001ApJ...557..616G}
{Gardner}, J.~P. 2001, \apj, 557, 616

\bibitem[{{Hahn} {et~al.}(2007{\natexlab{a}}){Hahn}, {Carollo}, {Porciani}, \&
  {Dekel}}]{2007MNRAS.381...41H}
{Hahn}, O., {Carollo}, C.~M., {Porciani}, C., \& {Dekel}, A.
  2007{\natexlab{a}}, \mnras, 381, 41

\bibitem[{{Hahn} {et~al.}(2007{\natexlab{b}}){Hahn}, {Porciani}, {Carollo}, \&
  {Dekel}}]{2007MNRAS.375..489H}
{Hahn}, O., {Porciani}, C., {Carollo}, C.~M., \& {Dekel}, A.
  2007{\natexlab{b}}, \mnras, 375, 489

\bibitem[{{Hahn} {et~al.}(2010){Hahn}, {Teyssier}, \&
  {Carollo}}]{2010MNRAS.405..274H}
{Hahn}, O., {Teyssier}, R., \& {Carollo}, C.~M. 2010, \mnras, 405, 274

\bibitem[{{Hatton} \& {Ninin}(2001)}]{2001MNRAS.322..576H}
{Hatton}, S. \& {Ninin}, S. 2001, \mnras, 322, 576

\bibitem[{{Heavens} {et~al.}(2000){Heavens}, {Refregier}, \&
  {Heymans}}]{2000MNRAS.319..649H}
{Heavens}, A., {Refregier}, A., \& {Heymans}, C. 2000, \mnras, 319, 649

\bibitem[{{Hoyle}(1949)}]{hoyle}
{Hoyle}, F. Problems of Cosmical Aerodynamics, ed. B.~J. M. \& V.~H.C. (Dayton,
  Ohio: Central Air Documents Office)

\bibitem[{{Jones} {et~al.}(2010){Jones}, {van de Weygaert}, \&
  {Arag{\'o}n-Calvo}}]{2010MNRAS.408..897J}
{Jones}, B.~J.~T., {van de Weygaert}, R., \& {Arag{\'o}n-Calvo}, M.~A. 2010,
  \mnras, 408, 897

\bibitem[{{Knebe} \& {Power}(2008)}]{2008ApJ...678..621K}
{Knebe}, A. \& {Power}, C. 2008, \apj, 678, 621

\bibitem[{{Lauberts}(1982)}]{1982euse.book.....L}
{Lauberts}, A. 1982, {ESO/Uppsala survey of the ESO(B) atlas}, ed. {Lauberts,
  A.}

\bibitem[{{Lee}(2011)}]{2011ApJ...732...99L}
{Lee}, J. 2011, \apj, 732, 99

\bibitem[{{Lee} \& {Erdogdu}(2007)}]{2007ApJ...671.1248L}
{Lee}, J. \& {Erdogdu}, P. 2007, \apj, 671, 1248

\bibitem[{{Lee} {et~al.}(2005){Lee}, {Kang}, \& {Jing}}]{2005ApJ...629L...5L}
{Lee}, J., {Kang}, X., \& {Jing}, Y.~P. 2005, \apjl, 629, L5

\bibitem[{{Lee} \& {Pen}(2000)}]{2000ApJ...532L...5L}
{Lee}, J. \& {Pen}, U. 2000, \apjl, 532, L5

\bibitem[{{Lee} \& {Pen}(2001)}]{2001astro.ph.11186L}
---. 2001, ArXiv Astrophysics e-prints

\bibitem[{{Lemson} \& {Kauffmann}(1999)}]{lemson}
{Lemson}, G. \& {Kauffmann}, G. 1999, \mnras, 302, 111

\bibitem[{{Maller} {et~al.}(2002){Maller}, {Dekel}, \&
  {Somerville}}]{2002MNRAS.329..423M}
{Maller}, A.~H., {Dekel}, A., \& {Somerville}, R. 2002, \mnras, 329, 423

\bibitem[{{Mu{\~n}oz-Cuartas} {et~al.}(2011){Mu{\~n}oz-Cuartas}, {Macci{\`o}},
  {Gottl{\"o}ber}, \& {Dutton}}]{2011MNRAS.411..584M}
{Mu{\~n}oz-Cuartas}, J.~C., {Macci{\`o}}, A.~V., {Gottl{\"o}ber}, S., \&
  {Dutton}, A.~A. 2011, \mnras, 411, 584

\bibitem[{{Navarro} {et~al.}(2004){Navarro}, {Abadi}, \&
  {Steinmetz}}]{2004ApJ...613L..41N}
{Navarro}, J.~F., {Abadi}, M.~G., \& {Steinmetz}, M. 2004, \apjl, 613, L41

\bibitem[{{Nilson}(1974)}]{1974UppOR...5....0N}
{Nilson}, P. 1974, Uppsala Astronomical Observatory Reports, 5, 0

\bibitem[{{Peebles}(1969)}]{peebles69}
{Peebles}, P.~J.~E. 1969, \apj, 155, 393

\bibitem[{{Pen} {et~al.}(2000){Pen}, {Lee}, \& {Seljak}}]{2000ApJ...543L.107P}
{Pen}, U.-L., {Lee}, J., \& {Seljak}, U. 2000, \apjl, 543, L107

\bibitem[{{Porciani} {et~al.}(2002){Porciani}, {Dekel}, \&
  {Hoffman}}]{2002MNRAS.332..325P}
{Porciani}, C., {Dekel}, A., \& {Hoffman}, Y. 2002, \mnras, 332, 325

\bibitem[{Schaap(2007)}]{SchaapThesis}
Schaap, W.~E. 2007, PhD thesis, University of Groningen

\bibitem[{{Schaap} \& {van de Weygaert}(2000)}]{DTFE}
{Schaap}, W.~E. \& {van de Weygaert}, R. 2000, \aap, 363, L29

\bibitem[{{Sharma} \& {Steinmetz}(2005)}]{2005ApJ...628...21S}
{Sharma}, S. \& {Steinmetz}, M. 2005, \apj, 628, 21

\bibitem[{{Slosar} {et~al.}(2009){Slosar}, {Land}, {Bamford}, {Lintott},
  {Andreescu}, {Murray}, {Nichol}, {Raddick}, {Schawinski}, {Szalay}, {Thomas},
  \& {Vandenberg}}]{2009MNRAS.392.1225S}
{Slosar}, A., {Land}, K., {Bamford}, S., {Lintott}, C., {Andreescu}, D.,
  {Murray}, P., {Nichol}, R., {Raddick}, M.~J., {Schawinski}, K., {Szalay}, A.,
  {Thomas}, D., \& {Vandenberg}, J. 2009, \mnras, 392, 1225

\bibitem[{{Sousbie} {et~al.}(2009){Sousbie}, {Colombi}, \&
  {Pichon}}]{2009MNRAS.393..457S}
{Sousbie}, T., {Colombi}, S., \& {Pichon}, C. 2009, \mnras, 393, 457

\bibitem[{{Springel}(2005)}]{gadget2}
{Springel}, V. 2005, \mnras, 364, 1105

\bibitem[{{Springel} {et~al.}(2005){Springel}, {White}, {Jenkins}, {Frenk},
  {Yoshida}, {Gao}, {Navarro}, {Thacker}, {Croton}, {Helly}, {Peacock}, {Cole},
  {Thomas}, {Couchman}, {Evrard}, {Colberg}, \& {Pearce}}]{2005Natur.435..629S}
{Springel}, V., {White}, S.~D.~M., {Jenkins}, A., {Frenk}, C.~S., {Yoshida},
  N., {Gao}, L., {Navarro}, J., {Thacker}, R., {Croton}, D., {Helly}, J.,
  {Peacock}, J.~A., {Cole}, S., {Thomas}, P., {Couchman}, H., {Evrard}, A.,
  {Colberg}, J., \& {Pearce}, F. 2005, \nat, 435, 629

\bibitem[{{Springel} {et~al.}(2001){Springel}, {White}, {Tormen}, \&
  {Kauffmann}}]{2001MNRAS.328..726S}
{Springel}, V., {White}, S.~D.~M., {Tormen}, G., \& {Kauffmann}, G. 2001,
  \mnras, 328, 726

\bibitem[{{van de Weygaert} \& {Schaap}(2007)}]{CosmicWeb}
{van de Weygaert}, R. \& {Schaap}, W. 2007, ArXiv e-prints

\bibitem[{{van den Bosch} {et~al.}(2003){van den Bosch}, {Abel}, \&
  {Hernquist}}]{2003MNRAS.346..177V}
{van den Bosch}, F.~C., {Abel}, T., \& {Hernquist}, L. 2003, \mnras, 346, 177

\bibitem[{{Vitvitska} {et~al.}(2002){Vitvitska}, {Klypin}, {Kravtsov},
  {Wechsler}, {Primack}, \& {Bullock}}]{2002ApJ...581..799V}
{Vitvitska}, M., {Klypin}, A.~A., {Kravtsov}, A.~V., {Wechsler}, R.~H.,
  {Primack}, J.~R., \& {Bullock}, J.~S. 2002, \apj, 581, 799

\bibitem[{{White}(1984)}]{white84}
{White}, S.~D.~M. 1984, \apj, 286, 38

\bibitem[{{Zel'Dovich}(1970)}]{Zel'Dovich}
{Zel'Dovich}, Y.~B. 1970, \aap, 5, 84

\bibitem[{{Zhang} {et~al.}(2009){Zhang}, {Yang}, {Faltenbacher}, {Springel},
  {Lin}, \& {Wang}}]{2009ApJ...706..747Z}
{Zhang}, Y., {Yang}, X., {Faltenbacher}, A., {Springel}, V., {Lin}, W., \&
  {Wang}, H. 2009, \apj, 706, 747

\end{thebibliography}

\end{document}